\documentclass{aastex631}
\usepackage{makecell}
\usepackage{subfigure}
\usepackage{threeparttable}
\usepackage{longtable}
\usepackage{booktabs}
\usepackage{tabularx}
\usepackage{CJK}
\usepackage{multirow}
\usepackage{hhline}
\usepackage{rotating}

\newcommand       \Teff         {T_{\rm {eff}}}
\newcommand       \feh          {\rm \left[Fe/H \right]}
\newcommand       \logg         {{\rm log}\ g}
\newcommand       \ab[2]        {{#1}_{\rm {#2}}}
\newcommand       \Gbp          {\ab{G}{BP}}
\newcommand       \K            {\ab{K}{S}}

\newcommand       \JK           {E_{J,\K}}
\newcommand       \rv           {R_{\rm {V}}}
\newcommand       \rbp          {R_{\rm {G_{BP}}}}
\defcitealias{Cao23}{Paper\,I}

\begin{document}
\begin{CJK*}{UTF8}{gbsn}

\title{Extinction of Taurus, Orion, Perseus and California Molecular Clouds Based on the LAMOST, 2MASS, and Gaia Surveys II: The Extinction Law}

\author[0000-0002-0316-1112]{ZheTai Cao (曹哲泰)}
\affiliation{Institute for Frontiers in Astronomy and Astrophysics,
            Beijing Normal University,  Beijing 102206, China}
\affiliation{School of Physics and Astronomy,
               Beijing Normal University,
               Beijing 100875, China}
\author[0000-0003-3168-2617]{Biwei Jiang (姜碧沩)}
\affiliation{Institute for Frontiers in Astronomy and Astrophysics,
            Beijing Normal University,  Beijing 102206, China}
\affiliation{School of Physics and Astronomy,
               Beijing Normal University,
               Beijing 100875, China}
\author[0000-0003-4489-9794]{Shu Wang (王舒)}
\affiliation{CAS Key Laboratory of Optical Astronomy, National Astronomical Observatories, Chinese Academy of Sciences, Beijing 100101, China}
\author[0000-0001-9328-4302]{Jun Li (李军)}
\affiliation{Center for Astrophysics, Guangzhou University, Guangzhou 510006, China}

\begin{abstract}
The extinction law from ultraviolet (UV) to infrared (IR) (0.2-24 $\mu$m) is determined by relying on the blue-edge method and color excess ratios for some nearby molecular clouds, from low mass star forming region to massive star forming region. The observational data are collected from nine photometric surveys, along with stellar parameters from the APOGEE and LAMOST spectroscopic surveys. Within the uncertainties, the optical ratio of selective to total extinction ($\rbp$) does not vary substantially across the clouds, irrespective of the density, specifically $\rbp =2.302\pm0.027$, where $\rbp \equiv A_{\Gbp}/E_{\Gbp,\ab{G}{RP}}$. The IR extinction law is consistent with \citet{Wang19_law}. The extinction law in the UV band is compromised by the shallow depth with $\ab{A}{V} \leq 2$ mag and is hard to describe by one parameter $R$. In addition, the extinction in the WISE/W1 band is significantly larger than in the Spitzer/IRAC1 band in the dense regions, which is attributed to the ice water absorption.
\end{abstract}


\section{Introduction} \label{sec:intro}

The extinction law, also known as extinction curve, can reflect the dust properties such as grain size, and is an important tool to study the environment of interstellar medium (ISM). The parameter $\rv$ is used to describe the shape of extinction curve \citep{F99,CCM89}, which is defined as $\rv\equiv A_{V}/E_{B,V}$ where $E_{B,V}$ is the color excess (CE) between the $B$ and $V$ band and $A_{\rm V}$ is the extinction in the $V$ band \citep{CCM89}. The value of $\rv$ is different towards various line of sights. \citet{Zhang2023_RvMap}, \citet{Schlafly2016_Rv} and \citet{Lee18_Rv} computed the 2D $\rv$ map on the Galactic scale, finding some distribution patterns along the Galactic longitude. Within the Taurus Molecular Cloud (TMC), \citet{Li2023_TMCLaw} studied the $\rv$ distribution deriving the average $\rv=3.13\pm0.32$. The reason why and how $\rv$ varies with environment is still controversial. Dust grains grow in dense environment \citep{Li2024_DustGrowth} and, if exposed to radiation, they are likely to be destroyed. As a result, it is reasonable to expect that there are some variation of extinction law with environment, from dense to diffuse ISM, or in different star forming regions. To investigate this problem, this work studies the extinction law in a wide wavelength range of three nearby molecular clouds (MCs) with different star forming activities.

The Taurus MC (TMC), Perseus MC (PMC), California MC (CMC) and Orion MC (OMC) are selected for this purpose. In our previous work \citep[hereafter \citetalias{Cao23}]{Cao23},  the 3D extinction maps towards these MCs are built, which separated CMC from TMC by the fact that CMC is at a distance of around 440-500pc apparently behind TMC at a distance of about 130-150pc. Since CMC is entirely behind TMC and much thinner than TMC, the extinction of the stars in this sightline is caused dominantly by TMC. To avoid confusion, CMC is dropped from the present study. Among the three target MCs, TMC is possibly the nearest and best-studied low mass star forming region, OMC is the closest massive star forming region, and PMC is an intermediate and low-mass star forming region. These three MCs not only have different activities, but also have both dense core and its surrounding diffuse region. In \citetalias{Cao23}, it is verified that some substructures include only one MC or its extinction is contributed mainly by one MC, which makes it possible to study the extinction law of a single MC instead of a complex combination of various types of MCs in one line of sight. Consequently, their extinction law can be compared to show the effect of environment on the extinction law.

The characteristics of extinction curve depends on the wavebands. In the UV band, the extinction curve shows significant variation with $\rv$ \citep{CCM89}. Besides, due to that the UV extinction is very severe, it is not possible to penetrate dense region in this band, while to diffuse region, the UV band is able to reveal the difference in $\rv$ effectively \citep{Sun18_UV_extinction}. In the optical band, the extinction curve shows some difference with $\rv$ although not so pronounced as in the UV band. To explore the UV-Optical extinction law, the stellar parameters are collected from LAMOST DR8 \citep{LMTDR8} data, and the photometric data are collected from GALEX \citep{GALEX}, XMM-OM \citep{XMM}, Swift/UVOT \citep{UVOT}, Gaia \citep{GaiaDR3}, Pan-STARRS \citep{PS1}, and SDSS \citep{SDSS16}. Furthermore, the extinction in the IR band is much weaker, and the stars with a large optical extinction can still be detectable in IR, which provides the possibility to study the extinction law in dense region. The stellar parameters are additionally taken from the near-IR spectroscopic survey APOGEE \citep{APGDR17} and the photometric data are taken from the 2MASS \citep{2MASS}, WISE \citep{WISE} and Spitzer surveys \citep{c2d,GLIMPSE,MIPSGAL,OrionSurvey,TaurusSurvey} to study the IR extinction law in dense and diffuse regions.

The paper is organized as follows, Section \ref{sec:data} describes the datasets, Section \ref{sec:BE} describes the blue-edge method used to determine the intrinsic color index of individual star, Section \ref{sec:CEs2MClaw} is about the method to derive the extinction law with the color excess, and Section \ref{sec:discussion} is the result and discussion.

\section{Data and Sample} \label{sec:data}

The intrinsic color indexes are the fundamental parameter to determine the color excess. Though there have been several works that determine the color index by stellar parameters and observed color index e.g. \citep{Jian2017_RevBE, Sun18_UV_extinction, Xue2016_IRlaw} in the way as we will work out here, still the color indexes are calculated independently in this work. One reason is that this work involves many more bands, in particular the UV bands, some of which are not dealt with previously. The other reason is that we try to improve the accuracy by a numerical method. The data to use includes two parts, stellar parameters from spectroscopy and brightness from photometry.

\subsection{Stellar Parameters}

As the extinction from UV to near-infrared (NIR) varies by a large amplitude, the data from multiple surveys are used to cover the extinction range. For the optical and UV bands, stellar parameters are taken from the LAMOST survey that is an optical stellar spectroscopic survey with a depth of about $G<17$ mag \citep{LMTDR8}. For the IR bands, they are from the APOGEE (Apache Point Observatory Galaxy Evolution Experiment) survey, which is a large-scale spectroscopic survey obtaining high-resolution and high signal-to-noise ratio spectra in the $1.51-1.70$ $\mu$m wavelength range \citep{APGDR17}. The LAMOST survey is able to detect the extinction to a depth of $A_{V} \approx 5$ \citepalias{Cao23} while the APOGEE survey is able to detect $A_{V} \approx 25-30$.

\begin{figure*}
    \centering
    \includegraphics{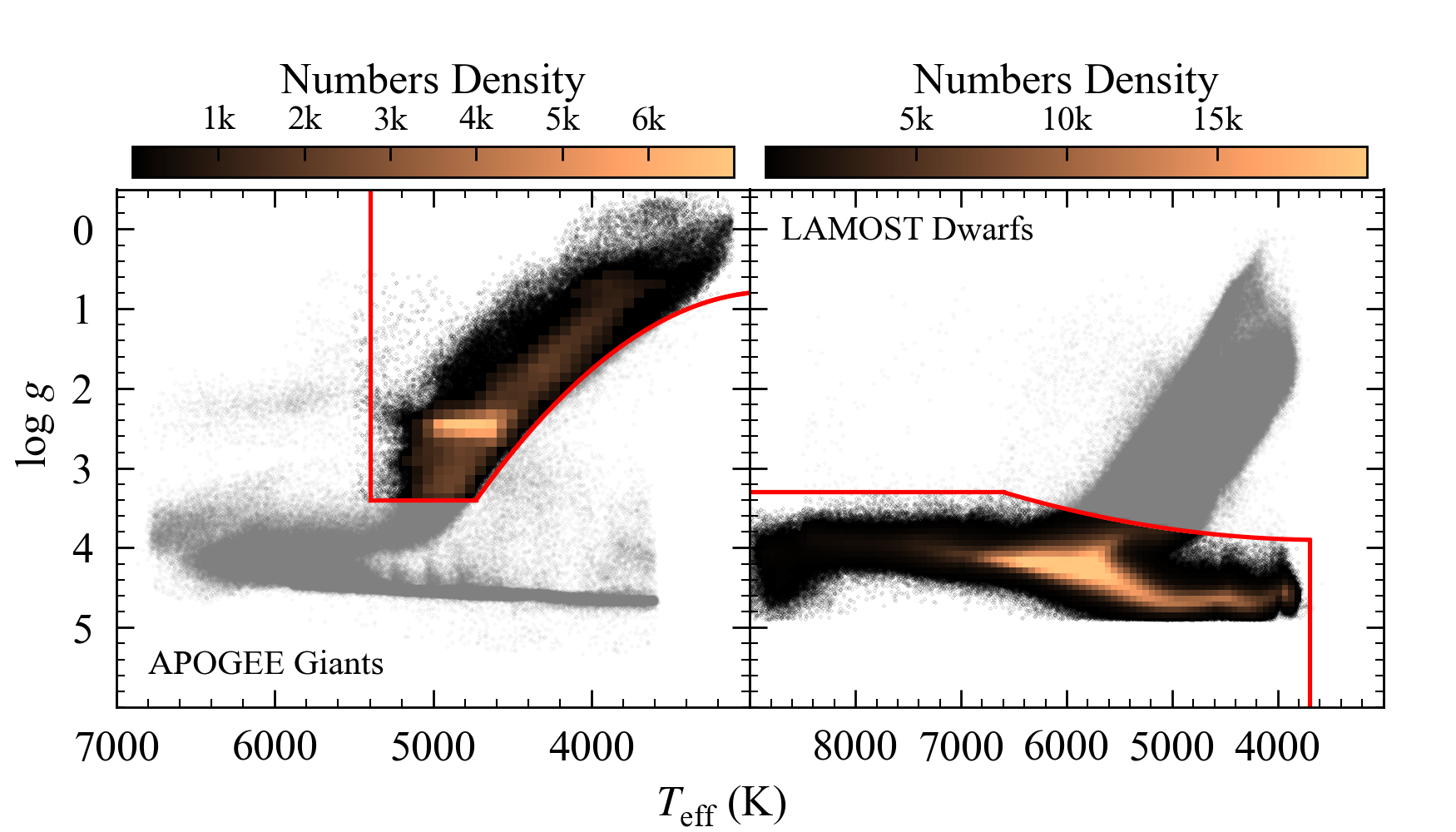}
    \caption{The Kiel diagram of the APOGEE (left) and LAMOST (right) stellar parameters. The black dots above the red line in the left panel and below the red line in the right panel are selected as giants and dwarfs, and the gray dots are dropped. The color indicates the number density.}
    \label{fig:DfsGts}
\end{figure*}

The LAMOST DR8 and APOGEE DR17 catalogs we use provide effective temperature ($\Teff$), surface gravity ($\logg$), and metallicity ($\feh$). After removing the duplicated sources by keeping the one with the highest signal-to-noise ratio, the following criteria are adopted to control the quality of the parameters:
\begin{enumerate}
\item $\feh$ is within [-1.0, 0.5].
\item The error of $\Teff$, $\feh$ and $\logg$ is smaller than 150K, 0.15 dex and 0.3 dex respectively.
\item The signal-to-noise ratio is larger than 10.
\item For the LAMOST survey, $3700\rm{K}<\textit{T}_{eff}<9000\rm{K}$, and for APOGEE, $3000\rm{K}<\textit{T}_{eff}<5500\rm{K}$
\end{enumerate}
About 4.7 million LAMOST stars and 0.64 million APOGEE stars are kept after this selection process. Moreover, only dwarfs in the LAMOST catalog are selected because: (1) the stellar parameters of dwarfs are more reliable, and (2) dwarfs significantly outnumber giants in LAMOST. Similarly, only giants are selected from the APOGEE catalog  because, (1) giants typically have a larger luminosity, which makes it still detectable even after a severe extinction; (2) the number of giants are much larger than dwarfs. The selection of the LAMOST dwarfs and the APOGEE giants is based on the Kiel diagram shown in Figure \ref{fig:DfsGts}, in principle according to surface gravity and effective temperature. Finally, a dataset with 3.8 million LAMOST dwarfs and 0.38 million APOGEE giants are obtained.

\subsection{Photometric Data}

For the photometric data, the GALEX, XMM-OM and Swift/UVOT surveys are used for UV, the SDSS, Gaia and Pan-STARRS surveys for optical, and the WISE, 2MASS and Spitzer surveys for IR. All the photometric catalogs are cross matched with the APOGEE for the IR bands or LAMOST dataset for the UV and optical bands within $1\arcsec$. The summary of photometric data is assembled in Table \ref{table:sources}.

\subsubsection{UV Photometric Data}\label{sec:UVdata}

\begin{enumerate}
\item The GALEX GR/6+7 \citep{GALEX} is the largest all-sky UV photometric database with high-quality. It has two filters, FUV (${\lambda_{\rm eff}}=152.8$nm) and NUV (${\lambda_{\rm eff}}=231.0$nm). The typical depth of FUV and NUV is 19.9 and 20.8 mag respectively. Here, we take the photometry $13.73<$FUV$<19.9$ and $13.85<$NUV$<20.8$ to exclude the possible saturation. The photometry error should be smaller than 0.5 mag.

\item XMM-SUSS \citep{XMM} is a catalogue of UV sources detected serendipitously by the Optical Monitor on-board the XMM-Newton observatory (XMM-OM, hereafter OM). The photometric data to use are from its three UV filters, i.e. UVW2 ($\ab{\lambda}{eff}=212.0$nm), UVM2 ($\ab{\lambda}{eff}=231.0$nm), and UVW1 ($\ab{\lambda}{eff}=291.0$nm), respectively. In comparison with GALEX, XMM-OM and Swift/UVOT are not all-sky survey projects and correspondingly not so frequently used.

SUSS 5.0 is a catalogue released in 2021 that contains 6 million sources, about 1 million of them have multiple observations. For single observations, all the data are kept directly, as for the sources with multiple observations, we prioritize selecting data with $quality\ flag=0$ or with the least photometric error if no observation with $quality\ flag=0$ is available. The limiting magnitudes of the three bands are UVW2$<20.2$, UVM2$<20.9$ and UVW1$<21.2$. The photometry error should be less than 0.5 mag.

\item Swift/UVOT Serendipitous Source Catalog \citep{UVOT} also provides photometry in six bands which is very similar to the OM module, i.e. UVW2 ($\ab{\lambda}{eff}=209.0$nm), UVM2 ($\ab{\lambda}{eff}=225.0$nm), UVW1 ($\ab{\lambda}{eff}=268.0$nm), and the classical UBV bands. Here, the first three UV bands are used.

The Swift/UVOT (hereafter UVOT) catalog contains about 6.2 million sources in total, and about 2 million have multiple entries in the source table due to repeated observations. Very similar to the XMM-OM catalog, we prioritize keeping the data with $quality\ flag=0$ or with the least photometric errors if no observation with $quality\ flag=0$ is available. In addition, the faint sources are dropped by limiting mag. It should be noticed that, different from OM data, the limiting magnitude differs for individual exposure. The faint sources are dropped by the limiting magnitude of its corresponding exposure. Roughly speaking, the limiting magnitude is around 20 for these three bands. The photometry error should be less than 0.5 mag.
\end{enumerate}

\subsubsection{Optical Photometric Data}\label{sec:Opticaldata}

\begin{enumerate}
\item The Gaia Data Release 3 (DR3) provides the best available photometry in optical, with unprecedented mmag precision photometry. In addition, it also provides parallaxes of $\mu$mas precision for entire celestial sphere. Based on parallax, \citet{BJdistance} estimate the distance of 1.47 billion stars. Gaia data are cross-matched with the LAMOST stellar parameters. The APOGEE stellar parameters are also associated with the Gaia photometry. As the Gaia catalog already contains the APOGEE stellar parameters, no new cross-match is performed in practice. The photometry error of Gaia is required to be less than 0.05 mag and only $\ab{G}{BP}$ and $\ab{G}{RP}$ band are used.

\item The Sloan Digital Sky Survey \citep[SDSS][]{SDSS16} provides photometry in five bandpasses, specifically, $\ab{u}{SDSS}$, $\ab{g}{SDSS}$, $\ab{r}{SDSS}$, $\ab{i}{SDSS}$ and $\ab{z}{SDSS}$ whose effective wavelength is about 357, 475, 620, 752 and 899 nm respectively. We take the photometric data from the latest data release DR16 \citep{SDSS16}. The criterion $clean=True$ is used to clean the sample. Considering saturation and sensitivity, it is required that the photometry magnitude should be $13<\ab{u}{SDSS}<22.3$, $14<\ab{g}{SDSS}<23.3$, $14<\ab{r}{SDSS}<23.1$, $14<\ab{i}{SDSS}<22.3$ and $12<\ab{z}{SDSS}<20.8$ \citep{SDSS16_maglim}. The photometry error should be less than 0.05 mag.

\item The Panoramic Survey Telescope and Rapid Response System 1 \citep[Pan-STARRS 1, hereafter PS1][]{PS1} observed the sky north of declination $-30\degr$. It provides the photometric data in five bands, $\ab{g}{PS1}$, $\ab{r}{PS1}$, $\ab{i}{PS1}$, $\ab{z}{PS1}$ and $\ab{y}{PS1}$ bands with the effective wavelength of about 490, 624, 756, 869 and 964 nm respectively. Here, we take the data with $14.5<\ab{g}{PS1}<22$, $15<\ab{r}{PS1}<21.8$, $15<\ab{i}{PS1}<21.5$, $14<\ab{z}{PS1}<20.9$ and $13<\ab{y}{PS1}<19.7$ with the photometry error less than 0.05 mag. Though PS1 share the $gri$ bands with SDSS, they are slightly different. So we treat them separately with different effective wavelength, and the consistency can be checked as well.
\end{enumerate}

\subsubsection{IR Photometric Data}\label{sec:IRdata}
\begin{enumerate}
\item The 2MASS project is a near-IR whole-sky survey \citep{2MASS}. The 2MASS/PSC contains photometric data in the $J$, $H$, and $\ab{K}{S}$ bands with the effective wavelength at about 1.24, 1.66, and 2.16 $\mu$m respectively. As the APOGEE catalog includes the 2MASS photometry data, no new cross-match is performed. The 2MASS photometry error is required to be less than 0.1 mag.

\item The Wide-field Infrared Survey Explorer (WISE) survey is a full-sky survey \citep{WISE}, offering photometry in four bands, i.e. W1, W2, W3 and W4 bands with the effective wavelength at 3.35, 4.60, 11.56 and 22.09 $\mu$m, respectively. The same as 2MASS, the APOGEE catalog already includes the WISE photometric data, no new cross-match is performed. The WISE photometry error is required to be  less than 0.1 mag.

\item The Spitzer Space Telescope is an IR space telescope. Here, we use the data from five bands in two instruments. The Infrared Array Camera (IRAC) has four bands centered at 3.6, 4.5, 5.8 and 8.0 $\mu$m \citep{IRAC} (Hereafter, $[3.6]$, $[4.5]$, $[5.8]$ and $[8.0]$) respectively. The other, the Multiband Imaging Photometer (MIPS), has three bands centered nominally at 24, 70, and 160 $\mu$m \citep{MIPS}, and we only use the 24 $\mu$m data (Hereafter $[24]$). The photometry error of all the bands is required to be smaller than 0.2 mag.

Because Spitzer is not an all-sky survey telescope, the data are collected from five large-scale survey projects: (1) Galactic Legacy Infrared Midplane Survey Extraordinaire \citep[GLIMPSE][]{GLIMPSE}, which focuses on the Galactic plane by using IRAC. (2) The Multiband Imaging Photometer for Spitzer Galactic Plane Survey (MIPSGAL) that focuses on the inner disk of the Milky Way using MIPS \citep{MIPSGAL}. (3) Spitzer From Molecular Cores to Planet-Forming Disks survey (C2D), which uses Spitzer/IRAC, MIPS and Infrared Spectrograph to observe various star-forming regions including the entire areas of five nearby large molecular clouds, though it mainly serves to supply the sources in PMC \citep{c2d,c2d_SFR}. (4) The Taurus Spitzer Legacy project, which uses IRAC and MIPS to map a region of 44 square degrees within the Taurus star-formation area \citep{TaurusSurvey}, and provides the aperture fluxes (and errors) for 2px, 5px, and 10px (corresponding to 2.44$\arcsec$, 6.1$\arcsec$, and 12.2$\arcsec$ respectively), here, we take the flux for 2 px as the flux\footnote{The reason we take flux for 2px is that the magnitude converted from fluxes for 2px is closest with C2D magnitude and has the least dispersion compared with 5px and 10px magnitude when we simply cross match the source in Taurus Spitzer Legacy project with C2D project.}. Finally, (5) Spitzer Orion Point Source Catalog \citep{OrionSurvey} to supplement sources in OMC. The Orion catalog has the IRAC and MIPS photometric results from 3.6 to 24 $\mu$m. All these projects are merged and then cross-matched with the APOGEE sources. In summary, the C2D, Taurus Spitzer Legacy Project and Spitzer Orion Point Source Catalog primarily cover the sources in PMC, TMC and OMC respectively.
\end{enumerate}

\section{The Intrinsic Color Index by the Blue-edge Method}\label{sec:BE}

A given set of stars with the same atmospheric parameters, i.e. $\Teff$, $\logg$ and $\feh$, should have the same intrinsic color index. The difference in the observed color index is then caused by reddening, which leads to the conclusion that the bluest star in the set is the one with minimal or no extinction. Based on this fact, the blue-edge method is used to determine the intrinsic color index by finding the bluest observed color index in the set that is at the blue edge in the diagram of $\Teff$ vs. observed color index. It was first suggested by \citet{Ducati_2001_BE} and then improved in recent studies (e.g. \citealt{W2014_IRlaw, Xue2016_IRlaw, wang24_jwstIR}).

\subsection{Division of Metallicity and Effective Temperature}

The route of the blue-edge method in this work is similar to the one in \citetalias{Cao23}. With the LAMOST stellar parameters, the intrinsic color index relative to $\Gbp$ is calculated as $\Gbp-\lambda$ or $\lambda-\Gbp$ for the UV and optical bands, while the APOGEE stellar parameters are used to calculate the intrinsic color indexes relative to $\K$ in the form of $\K-\lambda$ or $\lambda-\K$ involving only the IR bands. The effective temperature, as the primary factor to influence the intrinsic color index, is divided into the bins with a width of about 200K.

The metallicity is the second factor to influence the intrinsic color index through its effect on opacity. \citet{Jian2017_RevBE} found that this influence is minimal in the IR band, however, it becomes noticeable in the optical and more pronounced in the UV band caused by increasing absorption lines of metallic elements at the shorter wavelength \citep{Sun18_UV_extinction}. The division of metallicity into multiple bins are then necessary, in particular in the optical and UV bands. On the other hand, increasing the number of bins results in the decrease of source counts within each bin and consequently the uncertainty of the result. To match the diversity of wavelength and data, the bin-width of metallicity is not constant. For $\K-W3$, $\K-[3.6]$, $\K-[4.5]$ and $\K-[5.8]$, a bin size of 0.3 dex in $\feh$ is applied, and for $\K-W4$ and $\K-[24]$, a wider bin size of 0.75 dex is chosen due to the smaller number of sources in these bands and the insensitivity to metallicity. Though the number of stars in the UV bands is neither numerous as in the IR, a much smaller bin size of 0.15 dex in $\feh$ is adopted considering the significant influence of metallicity. For the optical and 2MASS/$JH\ab{K}{S}$ bands, a width of 0.15 dex is taken since there are enough sample stars.

\subsection{The Intrinsic Color Index}
\begin{figure*}
    \centering
    \subfigure[$\Gbp-\ab{G}{RP}$]{
        \includegraphics[width=0.48\linewidth, height=0.8\textheight]{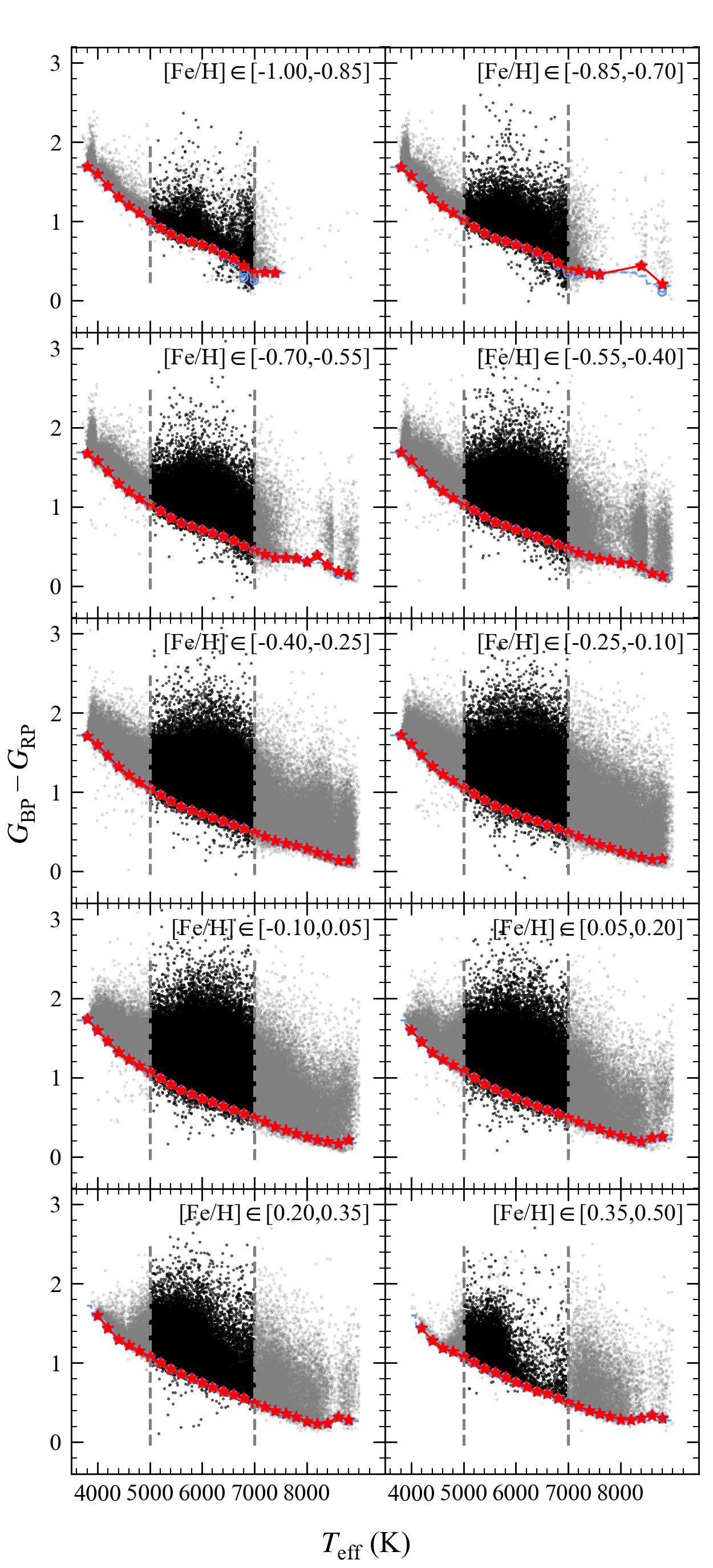}
        \label{fig:BE_Grp}
    }
     \hfill
    \subfigure[FUV-$\Gbp$]{
        \includegraphics[width=0.48\linewidth, height=0.8\textheight]{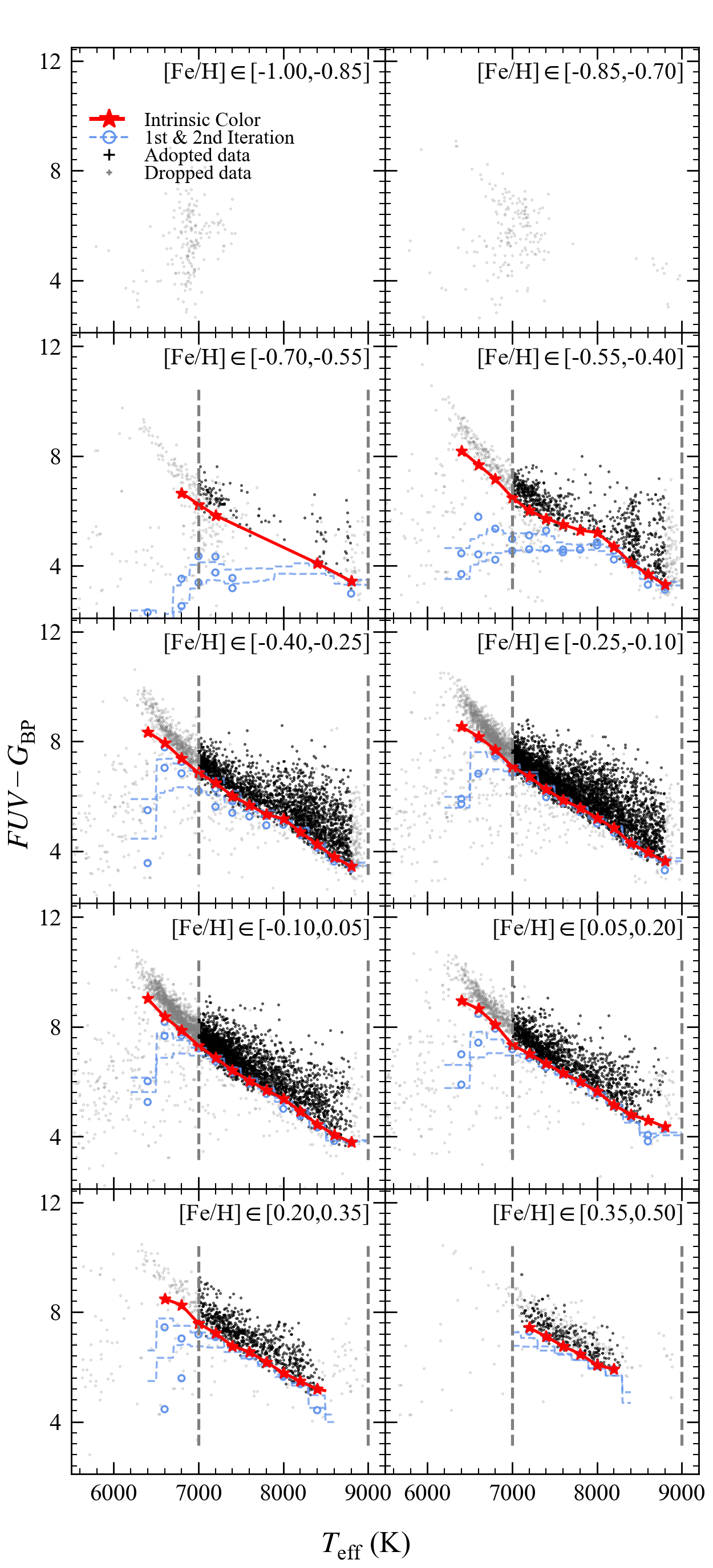}
        \label{fig:BE_FUV}
    }
    \caption{The blue-edge method used to determine the intrinsic color index for LAMOST dwarfs in different metallicity group. The dots denote the stars used in blue-edge method. The black dots are the final samples used to calculate extinction curve after quality control within a certain $\Teff$ range, which will be discussed later in Section \ref{sec:wls_shift} and Section \ref{sec:dropping_outliers}, the gray dots are the dropped ones. the red dots are the final `extinction-free' stars after cycling clipping, the blue dots are the results from Random Forest Regression in 1st and 2nd iteration.}
    \label{fig:BE_LMT}
\end{figure*}

Instead of selecting the bluest individual star, the bluest 5$\%$ stars in a bin are selected as the zero-extinction sources to reduce the effect of random error. In Figure \ref{fig:BE_Grp}, the points denote the observed $\Gbp-\ab{G}{RP}$ which has a sharp blue edge. The red points denote the bluest 5$\%$ `extinction-free' stars in each $\Teff$ bin. For some bins with a number of sources fewer than 20 or without a clear blue edge, as shown in the first two metallicity bins in Figure \ref{fig:BE_FUV}, the blue-edge method becomes unreliable and these bins are dropped in further calculation.

\begin{figure*}
    \centering
    \subfigure[{$\K-[3.6]$}]{
        \includegraphics[width=0.35\linewidth, height=0.9\textheight]{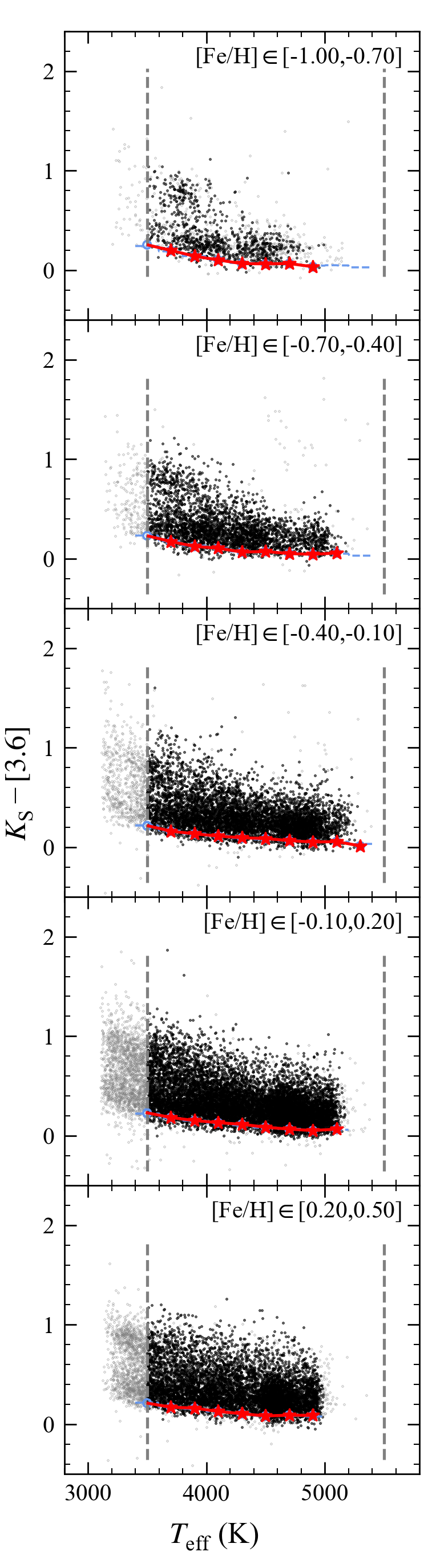}
        \label{fig:BE_IRAC36}
    }
     \hfill
    \subfigure[$\K-W4$]{
        \includegraphics[width=0.35\linewidth, height=0.45\textheight]{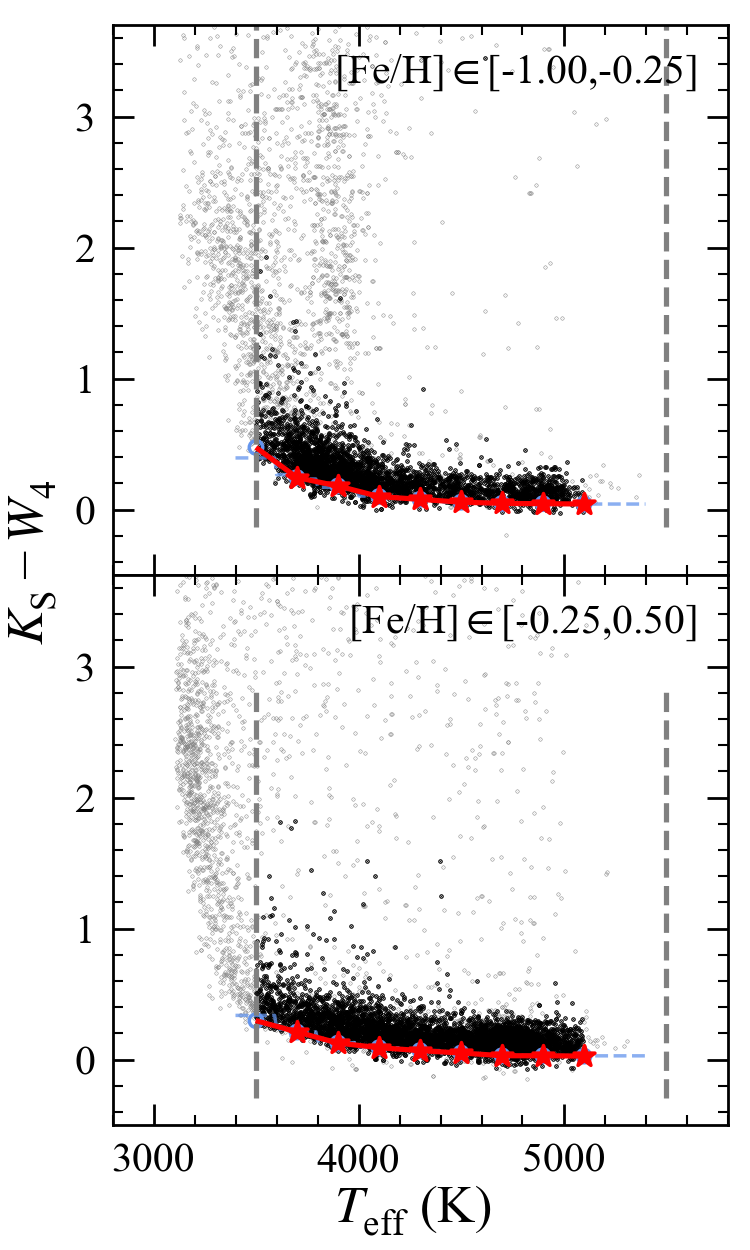}
        \label{fig:BE_W4}
    }
    \caption{The blue-edge method used to determine the intrinsic color index for APOGEE giants. It is similar to Figure \ref{fig:BE_LMT}, but the samples are from APOGEE giants.}
    \label{fig:BE_APG}
\end{figure*}

In the UV band, it occurs that some of the observed color index is bluer than the expected blue edge. The situation of FUV$-\Gbp$ is the most evident in this aspect, where the blue-ward shift cannot be fully accounted by the photometric error. As displayed in Figure \ref{fig:BE_FUV}, the FUV excess is apparent that increases towards low $\Teff$, which is explained by the increasing UV emission of chromospheric activity in low temperature stars \citep{Sun18_UV_extinction}. But these stars are not `extinction-free' stars, and they are removed by cycling clipping. Specifically, after the 5$\%$ bluest stars in each grid are selected in the first time, Random Forest Regression (RFR) \footnote{A RFR is a meta estimator that fits a number of decision trees to the data set and uses average values to improve the predictive accuracy, while controls overall fitting. \citet{Chen19_dustmap} uses RFR to estimate the extinction.}, one of the most effective machine-learning models for predictive analysis is applied to fit the relationship between intrinsic color index versus $\Teff$ and $\feh$. The stars that satisfy the following condition are removed:
\begin{equation}\label{equ:BE_sigmaclip}
{\rm ColorIndex_{observed} < IntrinsicColor_{predicted}-3\sigma_{CI}}
\end{equation}
The uncertainty of the intrinsic color index comes from both photometric error and the blue-edge method. The mean
error of the intrinsic color index is calculated by $\sigma_{\rm CI}=\sqrt{\sigma_{\lambda_{1}}^2+\sigma_{\lambda_{2}}^2+\sigma_{\rm BlueEdge}^{2}}$, where $\sigma_{\lambda_{1}}$ and $\sigma_{\lambda_{2}}$ represent the photometry error, and the error induced by the bluest edge $\sigma_{\rm BlueEdge}\approx30$ mmag \citep{Jian2017_RevBE}. Combining these three parts, $\sigma_{\rm{CI}}$ is shown in the `$\sigma_{\lambda_{1}\lambda_{2}}$' column in Table \ref{table:sources}. Then, the 5$\%$ bluest stars are selected for the second time, and the outliers are again removed by Equation \ref{equ:BE_sigmaclip}  until no star be removed in next step. Finally the real `extinction-free' stars can be selected. Figure \ref{fig:BE_FUV} illustrates this process, where the red dots are the final predictions of extinction-free stars after cycling clipping, and the blue dots are the results in the first two iterations. The cycling clipping method is highly effective in removing outliers in some band such as FUV$-\Gbp$ or $\Gbp-\ab{i}{SDSS}$. For most of the circumstance, such as in Figure \ref{fig:BE_Grp}, \ref{fig:BE_IRAC36}, \ref{fig:BE_W4}, almost no stars are removed, so the red dots are close to the blue dots.

The relation between the intrinsic color index and $\Teff$ and $\feh$ is so complex that it is hard to fit the relation with an analytic function. The \texttt{LinearNDInterpolator} from PYTHON package \texttt{SCIPY} \citep{Scipy} provides a method to perform an interpolation with the grid points in 2D. Here, it is used to fit the relation of the intrinsic color index with $\Teff$ and $\feh$. The reason we do not use RFR to fit the intrinsic color index is that the regression results from RFR is a `step' surface, instead, the results from \texttt{LinearNDInterpolator} is more smooth and continuous.

\section{Extinction Law of Molecular Clouds}\label{sec:CEs2MClaw}

\subsection{The Targeted Molecular Cloud Substructures Sample}\label{sec:CE_MCsSample}

\begin{figure*}
    \centering
    \includegraphics[width=\linewidth]{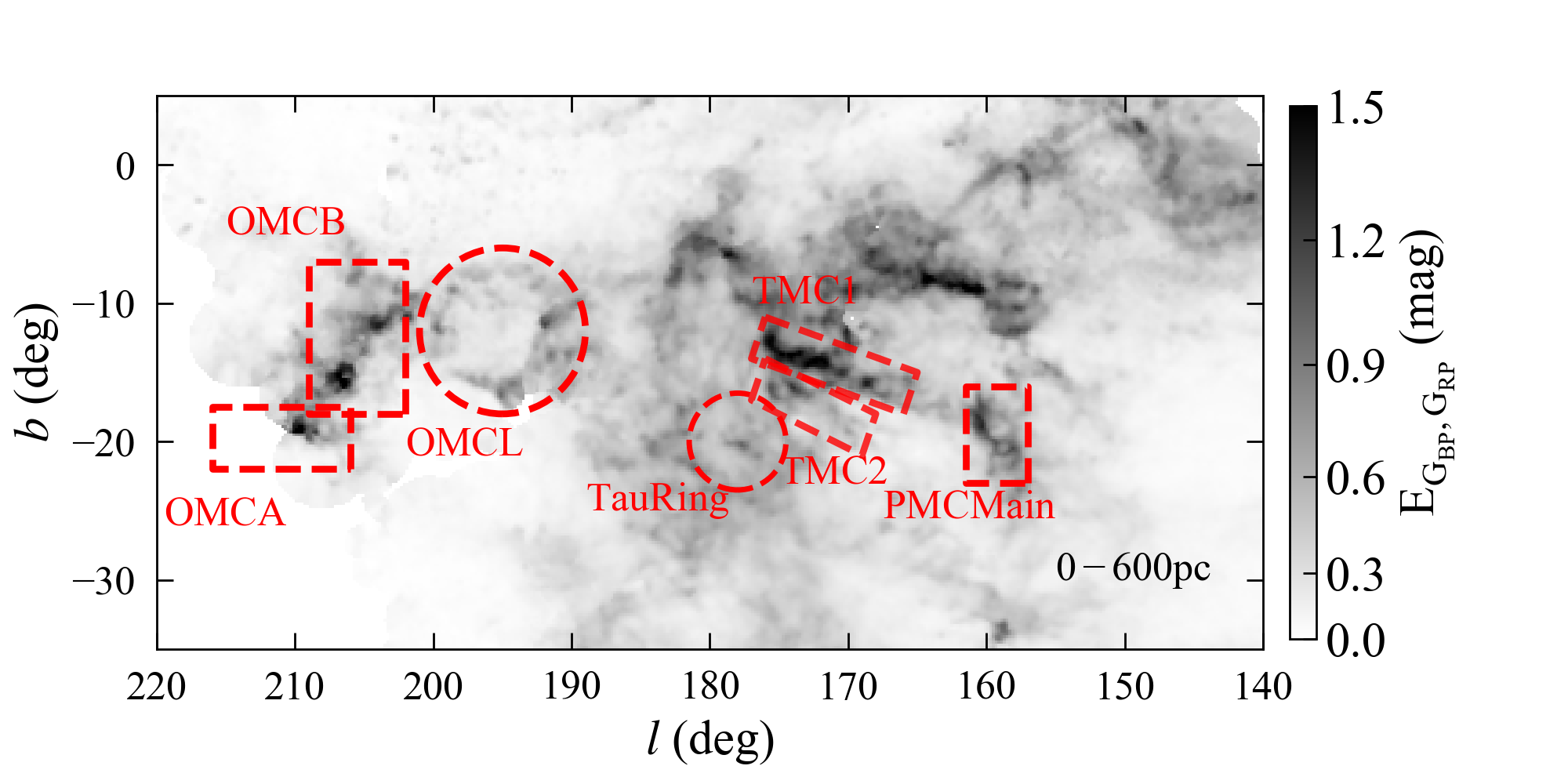}
    \caption{The extinction map of the studied regions integrated to a distance of 600 pc from \citetalias{Cao23}. The dashed red line outlines the region of targeted MC substructures sample.}
    \label{fig:dustmap}
\end{figure*}

A large structure variety exists even in one molecular cloud. For example, TMC has a few substructures like TMC1, TMC2, Tau Ring and filament that span a range of extinction from $E_{\Gbp,\ab{G}{RP}}$ of about 0.5 mag to 1.5 mag \citepalias{Cao23}. To study the influence of star-forming activity and interstellar environment on the extinction law, some typical substructures from the three MCs are selected other than the whole MC as an entity. Based on the 3D extinction map of a large area covering all the three MCs and the analysis of their substructures, the following sub-clouds are selected: TMC1, TMC2 from TMC, Orion A, Orion B and $\lambda$ Orionis from OMC (hereafter, OMCA, OMCB and OMCL), and Perseus Main from Perseus MCs (hereafter PMCMain). Additionally, TauRing in TMC is supplemented to represent the diffuse environment. The samples location are the same as listed in Table 2 of \citetalias{Cao23} except OMCL and TauRing. OMCL is a cone region centered at $(l,b) = (195\degr, -12\degr)$ with a larger radius of $6\degr$ compared with $5\degr$ in \citetalias{Cao23}, and TauRing is centered at $(l,b) = (178\degr, -20\degr)$ with a radius of $3.5\degr$. All the samples are outlined with red line in Figure \ref{fig:dustmap}.

These seven substructures are selected not only because they represent various radiation field and environment, but also because that the extinction towards these region sightlines is mostly contributed by the studied substructures instead of other foreground or background clouds according to the 3D dustmap in \citetalias{Cao23}, which may benefit from their location slightly deviating from the Galactic plane. In addition, almost all the stars that trace the extinction are behind the MCs since all the seven substructures are very close with a distance all less than 500pc and some less than 300pc.

\subsection{From Color-Excess Ratio to Relative Extinction}

Color-Excess Ratio (CER) is a widely used parameter to represent the extinction law based on the CE-CE diagram. \citet{Wang19_law} studied the optical to MIR extinction law and \citet{Meingast2018_OrionLaw} and \citet{Li2023_TMCLaw} studied extinction law of Orion and Taurus MC with the CER method. From the CER to the relative extinction $A_{\lambda}/A_{\ab{\lambda}{ref}}$, a conversion must be performed. In details, the CER $k_{\lambda}$ of two CEs is:
\begin{equation}\label{equ:CERs}
k_{\lambda} = \frac{E(\lambda-\ab{\lambda}{ref})}{E(\ab{\lambda}{com}-\ab{\lambda}{ref})} = \frac{A_{\lambda}-A_{\ab{\lambda}{ref}}}{A_{\ab{\lambda}{com}}-A_{\ab{\lambda}{ref}}}
\end{equation}
where $A_{\lambda}$ is the extinction in the $\lambda$ band, $\ab{\lambda}{ref}$ is the reference band and $\ab{\lambda}{com}$ is the comparison band. The extinction in the reference and comparison band should be measured as accurate as possible. For optical and UV band, we use $\Gbp$ and $\ab{G}{RP}$ respectively, since Gaia typically has a precise photometry as well as a full-sky coverage. For IR band, 2MASS bands are commonly used, here, $\K$ and $J$ are used as reference and comparison band respectively. Equation \ref{equ:CERs} can be manipulated to a different form as following:
\begin{equation}\label{equ:CER2ext}
A_{\lambda}/A_{\ab{\lambda}{ref}} = 1+k_{\lambda}(A_{\ab{\lambda}{com}}/A_{\ab{\lambda}{ref}}-1)
\end{equation}
The calculation of relative extinction $A_{\lambda}/A_{\ab{\lambda}{ref}}$ thus requires a conversion factor, $A_{\ab{\lambda}{com}}/A_{\ab{\lambda}{ref}}$. The value of $A_{\ab{\lambda}{com}}/A_{\ab{\lambda}{ref}}$ influences the result $A_{\lambda}/A_{\ab{\lambda}{ref}}$ and consequently the extinction law. However, it does not change the relative difference of the extinction curve since the same value of $A_{\ab{\lambda}{com}}/A_{\ab{\lambda}{ref}}$ is applied to all the bands. The adopted relative extinction is $A_{\ab{G}{RP}}/A_{\Gbp}=0.588$ from \citet{Wang19_law} and $A_{J}/A_{\K}=2.88$ from \citet{W2014_IRlaw}. 

To simplify the calculation of relative extinction, CEs are manipulated into a uniform format. For UV and optical band, CE is expressed as $E(\Gbp-\lambda)$, and for IR band, CE is $E(\lambda-\K)$. The relative extinction $A_{\lambda}/A_{\Gbp}$ or $A_{\lambda}/A_{\K}$ is then calculated as following:
\begin{equation}\label{equ:AxAlambda}
\begin{array}{ll}
\frac{A_{\lambda}}{A_{\Gbp}}= 1+k_{\lambda}(\frac{A_{\ab{G}{RP}}}{A_{\Gbp}}-1)\\
\frac{A_{\lambda}}{A_{\K}}= 1+k_{\lambda}(\frac{A_{J}}{A_{\K}}-1)
\end{array}
\end{equation}

\subsection{Influence of Stellar Parameters and Extinction on Color-excess-ratios }\label{sec:wls_shift}

Before calculating the average extinction law, the influence of extinction and stellar parameters on the CER should be addressed. \citet{Wang19_law} highlighted the curvature of the CE-CE diagram, which varies in accordance with the extinction. The wide bands such as the Gaia $G$ band or the UV bands are more likely to exhibit a large curvature, while the IR bands almost exhibit no curvature. \citet{Foster2013_PMClaw} pointed out that the CE-CE diagram exhibits a curve rather than a straight vector as extinction increases (as shown in Figure 6 in \citet{Foster2013_PMClaw}). However, the curvature of the CE-CE diagram can be ignored if the extinction is within a few magnitudes, in other words, the curvature of CE-CE diagram can be negligible in diffuse region or in the IR band. The other factor is stellar parameters to affect the CER. \citet{Zhang2023_CERs} studied the correlation between extinction coefficient and temperature and \citet{Sun18_UV_extinction} studied the variation of CER with extinction and $\Teff$. They both found that the CER changes significantly with $\Teff$.

To investigate the influence of stellar parameters and extinction on different bands, we adopt a similar method as in \citet{Sun2021_UVmap},  by considering the change of effective wavelength ($\ab{\lambda}{eff}$) of the filters. The definition of $\ab{\lambda}{eff}$ is:
\begin{equation}\label{equ:lambda_eff}
\ab{\lambda}{eff}=\frac{\int \lambda F(\lambda)S(\lambda)R(\lambda)d\lambda}{\int F(\lambda)S(\lambda)R(\lambda)d\lambda}
\end{equation}
where $F(\lambda)$ is the intrinsic stellar SED, which is taken from the ATLAS9 model for the dwarf stars with $\Teff$ between 4250K to 8500K, $\logg = 4.5$ and $\feh=0$. $S(\lambda)$ is the filter transmission, and $R(\lambda)$ is the extinction curve which is taken from \citet{F99} with $\rv=3.1$ at five extinctions that is $E_{\Gbp,\ab{G}{RP}}\in[0,0.25,0.5,0.75,1]$ mag ($A_{\rm V} \sim 2.2 E_{\Gbp,\ab{G}{RP}}$).

\begin{figure*}
    \centering
    \includegraphics{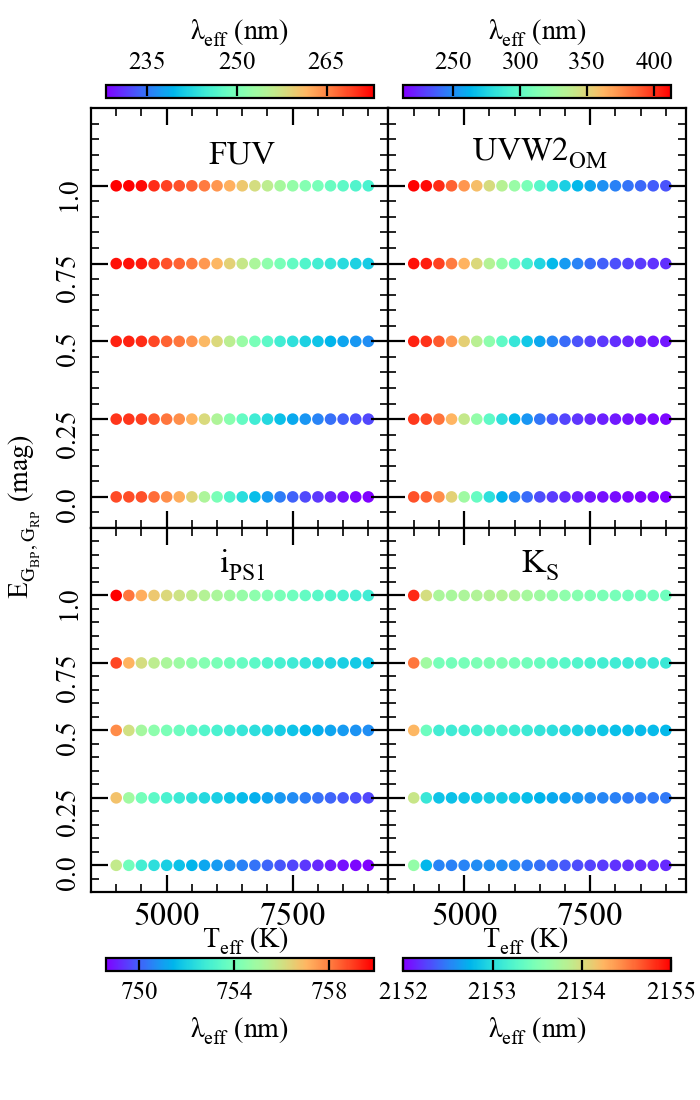}
    \caption{The shift of effective wavelength denoted by color with $T_{\rm eff}$ (x-axis) and extinction (y-axis) in four filters. }
    \label{fig:wls}
\end{figure*}

Some typical results of effective wavelength shift with $\Teff$ and extinction are displayed in Figure \ref{fig:wls}. The shift is mainly caused by $\Teff$. In comparison with optical and IR bands, the UV band is greatly shifted, especially in the OM and UVOT bands. The reason is the existence of `red-leaks' in the filter transmission curve \citep{UVOT} that allow longer wavelength photons passing through the filter. The UV flux of low temperature stars is very small so that a slight leak may lead to a significant increase of flux by the longer wavelength photons. In some extreme circumstance, such as the $UVW1_{\rm UVOT}$, $UVM2_{\rm UVOT}$ and $UVW2_{\rm UVOT}$ bands, the $\ab{\lambda}{eff}$ under the Vega spectrum is 209nm, 225nm and 268nm, but it becomes 293nm, 240nm and 314nm if under the SED of a 6000K dwarf.

Since the maximum extinction of the LAMOST sample dwarfs is about $E_{\Gbp,\ab{G}{RP}}\approx 2$ mag in the studied optical bands and $E_{\Gbp,\ab{G}{RP}}\approx 1$ mag in the UV bands, the influence of extinction can be ignored, and the influence on the IR bands can also be ignored for the APOGEE sample giants. However, the $\Teff$ effect can be significant. In order to reduce such effect, a more strict cutoff is applied to $\Teff$ in the following analysis. For the FUV bands, $\Teff$ is limited in $[7000\rm{K}, 9000\rm{K}]$ of the LAMOST catalog, and for other bands $\Teff$ is limited to $[5000\rm{K},7000\rm{K}]$ for the LAMOST catalog. Since the wavelength shift is negligible in the IR band, the $\Teff$ range for APOGEE dataset remains the same, i.e. $\Teff \in [3500\rm{K},5500\rm{K}]$.

\subsection{The Outliers}\label{sec:dropping_outliers}

In the bands such as W3, W4 and $[24]$, some stars exhibit IR excess that they are outliers in corresponding CE-CE diagram which need to be excluded. Here, we filter the data by sigma clipping in all the CE-CE diagram. For each CE-CE diagram, the data is binned with an interval of 0.1 mag in $E_{\Gbp,\ab{G}{RP}}$ or $E_{J,\K}$, then a 3-sigma cycling clipping is performed in each bin to drop the outliers.

\begin{figure*}
    \centering
    \subfigure[CERs with LAMOST dataset]{
        \includegraphics[width=0.9\textwidth, height=0.43\textheight]{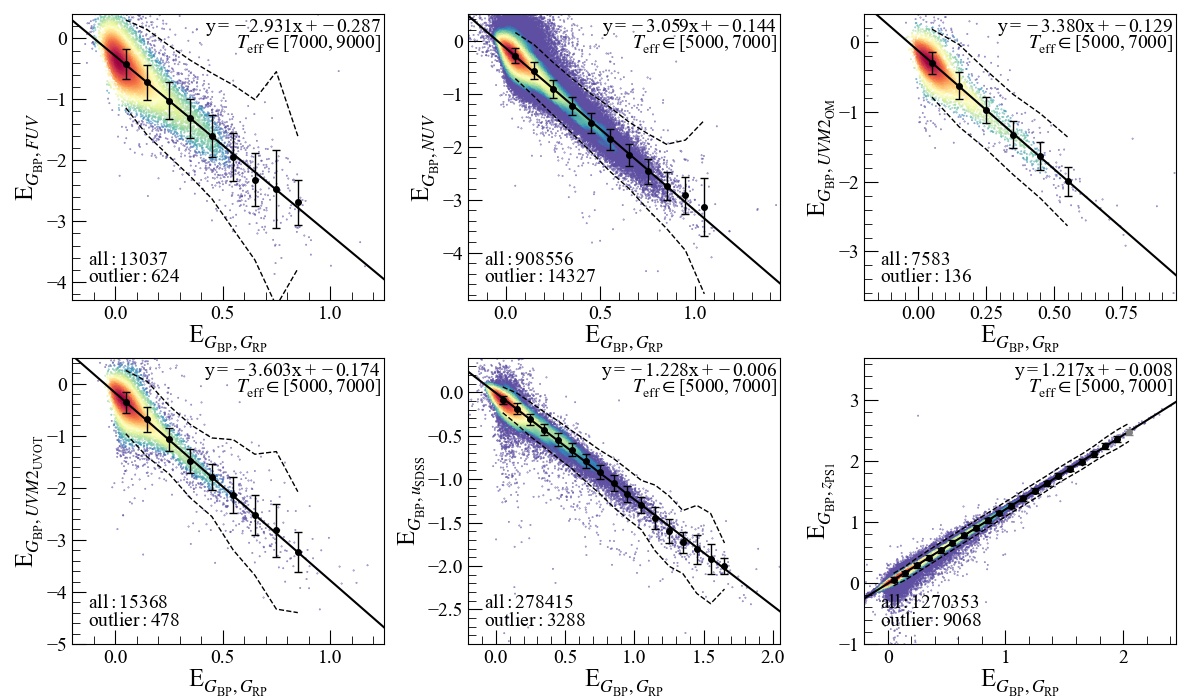}
        \label{fig:CERs_LMT}
    }
    \hfill
    \subfigure[CERs with APOGEE dataset]{
        \includegraphics[width=0.85\textwidth, height=0.43\textheight]{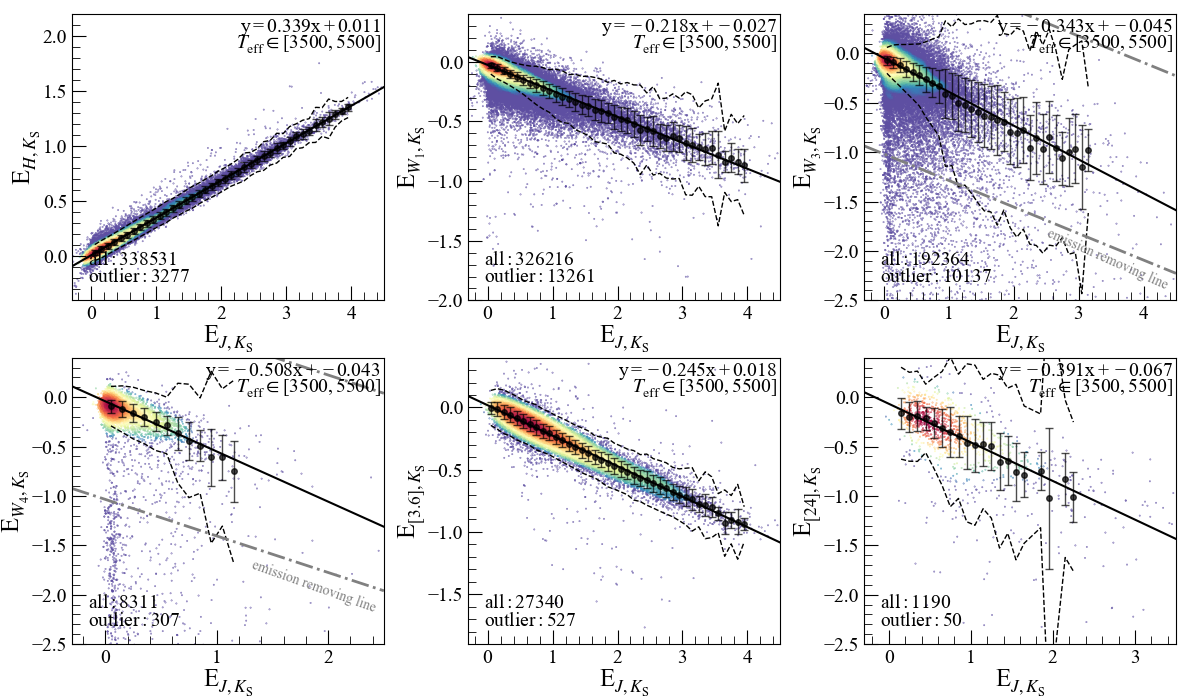}
        \label{fig:CERs_APG}
    }
    \caption{The color excess vs. color excess diagram for the whole sample, with the color denoting the star number density. The black dashed lines are the 3 sigma boundary, stars outside of which are dropped. For W3 and W4 band, as shown by the gray line in Figure \ref{fig:CERs_APG}, a dot dashed line is applied to roughly remove the silicate emission stars. The vertical black bar is the mean and standard deviation of the color excess within the bin, and the thick black line is the fitting result. The number of outliers is marked on the bottom left, the fitting result and the $\Teff$ of the stars are marked on the top right.}
    \label{fig:CERs}
\end{figure*}

The CE-CE diagram for the entire dataset of the band is shown in Figure \ref{fig:CERs}, where the color denotes the density of stars, the black bar indicates the mean and standard deviation of the data,  and the stars beyond 3 sigma indicated by the black dashed line are removed. For the bands of PS1, SDSS, 2MASS and some of the Spitzer, the linear relation in the CE-CE diagram is very tight and only few stars are removed. On the other hand, there are a significant number of outliers in the  $E_{W3,\K}$ and $E_{W4,\K}$ vs. $E_{J,\K}$ diagrams. The outliers are too many so that even the 3-sigma clipping cannot fully remove them, therefore, a pre-selection is performed for the W3 and W4 band. \citet{Xue2016_IRlaw} calculated the linear relation averaged over the all-sky between $E_{W3,\K}$ and $E_{W4,\K}$ versus $E_{J,\K}$, that is $E_{W3,\K}=-0.269*E_{J,\K}+0.016$ and $E_{W4,\K}=-0.370*E_{J,\K}+0.036$. On this basis, a range of $\pm1$ is added in the $y$-axis so that the stars beyond this range is removed in advance. It should be noted that the Spitzer/[24] band may be affected by the circumstellar silicate emission as well as the WISE/W3 and W4 band, but the outliers in this band is not so numerous that no additional pre-selection is required.

\subsection{The linear fitting between color excesses}\label{sec:Extinction_Curve}

\begin{figure*}
    \centering
    \includegraphics[width=\textwidth]{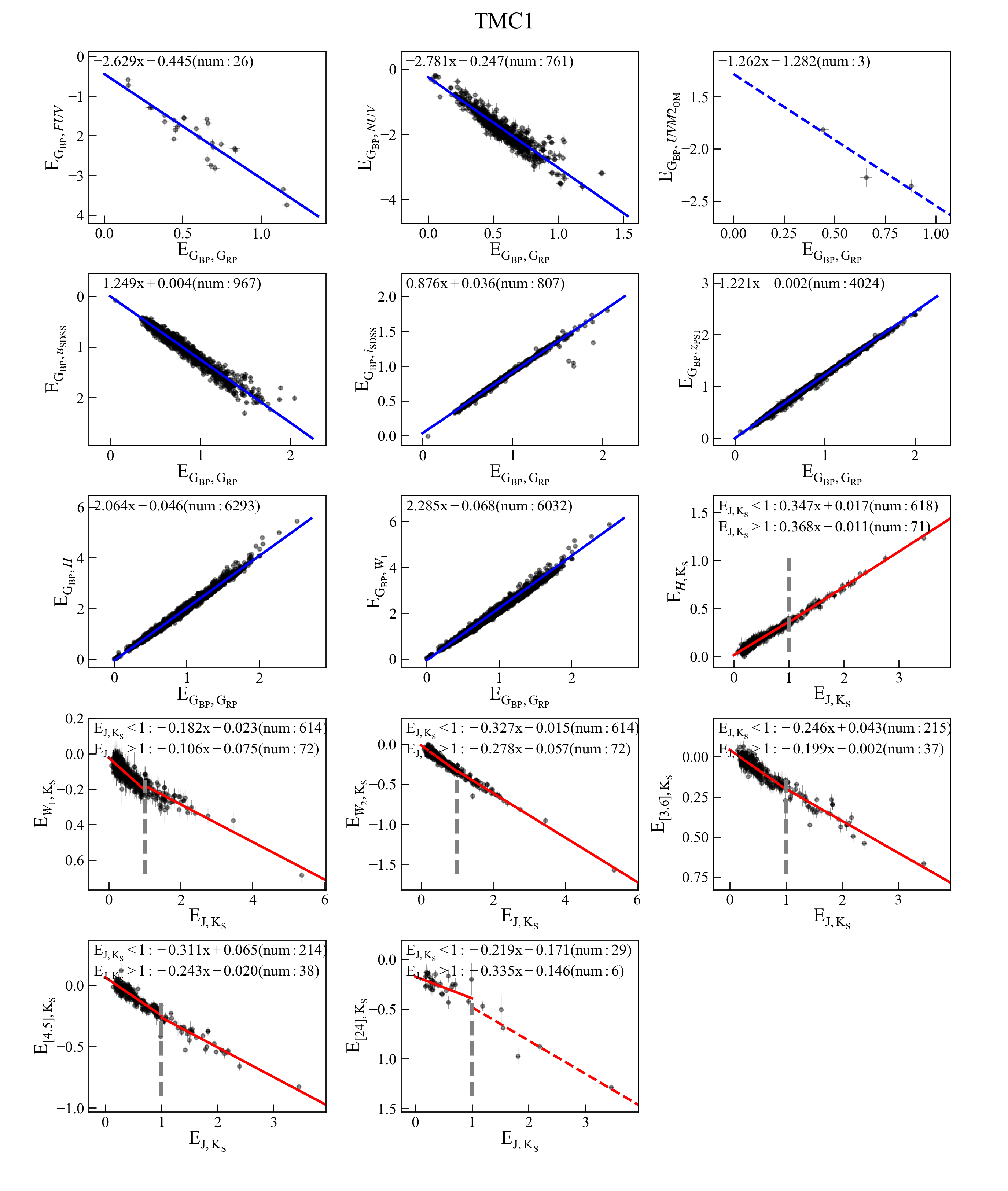}
    \caption{The linear fitting of the CE-CE diagram for the TMC1 sub-cloud. The black dots are the stars in the MC regions, the blue and red lines are the fitted line, in the UV-optical and IR band respectively. For the IR bands, the fitting is performed in two parts: diffuse region where $E_{J,\K}<1$ mag and dense region where $E_{J,\K}>1$ mag, separated by the gray dashed line. The dashed line indicates that the number of stars in the sample is fewer than 10 so that the result may be unreliable. The results of fitting and the number of stars used for fitting are listed on the top left in each panel.}
    \label{fig:slope_TMC1}
\end{figure*}

\begin{figure*}
    \centering
    \includegraphics[width=\textwidth]{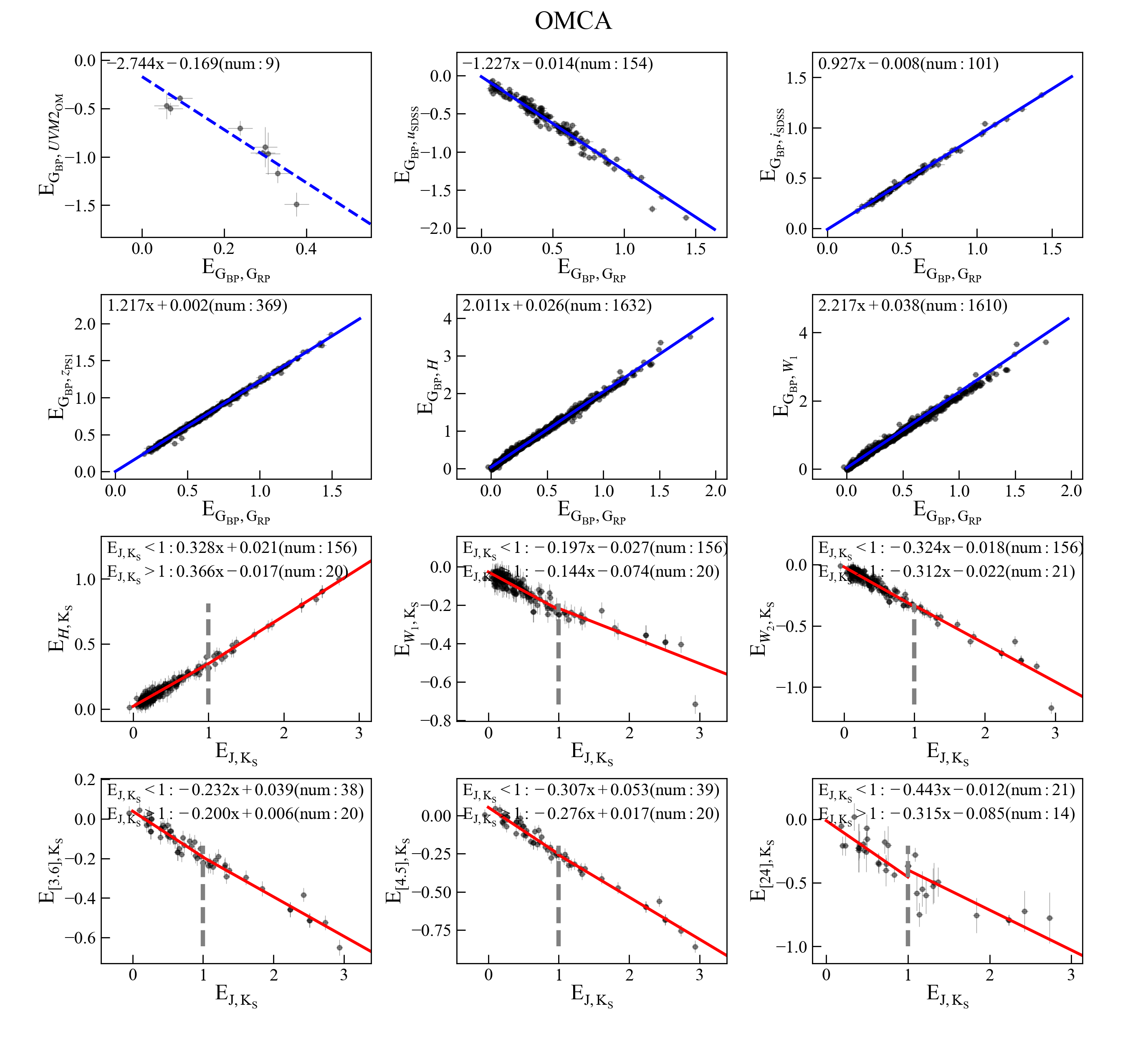}
    \caption{The same as Figure \ref{fig:slope_TMC1}, but for OMCA. Due to the lack of enough tracers, some bands are missed here in comparison with the TMC1 case due to lack of data.}
    \label{fig:slope_OMCA}
\end{figure*}

With the selected data by the method described above, the color excess ratio is determined by fitting the slope between $E(\Gbp-\lambda)$ and $E_{\Gbp,\ab{G}{RP}}$ or $E(\lambda-\K)$ and $E_{J,\K}$ via the Least Square Method in each band with the stars in the selected substructure regions of the three MCs, and the fitting error is expressed by the dispersion of the residual of the linear fit. In IR bands, the CERs are fitted separately in diffuse region ($E_{J,\K}<1$ mag) and dense region ($E_{J,\K}>1$ mag). Figure \ref{fig:slope_TMC1} and Figure \ref{fig:slope_OMCA} show the linear fitting result in TMC1 and PMCMain as examples. All the fitting results are listed in Table \ref{table:curve}.

\section{Result and Discussion}\label{sec:discussion}

The derived extinction curves expressed as color excess ratio vs. wavebands are displayed in Figure \ref{fig:law_UVoptical} and \ref{fig:law_IR} for the sample of the selected substructures of the MCs. The $\ab{\lambda}{eff}$ is calculated using the method described in Section \ref{sec:wls_shift}, with the central value calculated at the median effective wavelength $\Teff$, and the horizontal bar representing $\ab{\lambda}{eff}$ at the lower and upper limits of $\Teff$ respectively. Apart from the wavebands of the OM, GALEX and UVOT instruments, the effective wavelength shift is insignificant. Due to different depth of extinction detected in the LAMOST and APOGEE survey, the extinction curve is analyzed from FUV to W2 band with the LAMOST data and J to $[24]$ with the APOGEE data. Moreover, the IR analysis is further carried out for dense ($E_{J,\K}>1$) and diffuse ($E_{J,\K}<1$) region.

\subsection{The UV-Optical-NIR Extinction Law from the LAMOST Data}\label{sec:discussion:lamost}

Within the LAMOST data, the maximum extinction reaches to $E_{\Gbp,\ab{G}{RP}} \approx 2$ in optical band and $E_{\Gbp,\ab{G}{RP}} \approx 1$ in UV band. With $A_{\rm V} \approx 2.2E_{\Gbp,\ab{G}{RP}}$ \citep{Sun2021_UVmap}, this converts to $A_{\rm V} \approx $4.4 or 2.2 mag, implying that the optical and UV sample stars only trace the translucent and diffuse part of the cloud respectively.

The global characteristics of the extinction curve from UV to NIR is most widely expressed by the $\rv$ value. Here it is determined from the CER with the F99 \citep{F99} extinction model. F99 model \citep{F99} is a popular $\rv$ dependent extinction model for the wavelength range from 0.1$\mu m$ to 3.4$\mu m$. By applying minimum chi-square estimation with data from GALEX/FUV to WISE/W2, we obtain a fitted value for $\rv$, i.e. $\rv^{\rm fit}$ for one parameter fitting, where 68$\%$ confidence interval is $\chi^{2}(\theta_{68\%})=\sqrt{\chi^{2}(\theta_{minium})+1}$ and the fitting error is calculated as $\sigma=\theta_{68\%}-\theta_{minium}$. Finally, the results are 3.094$\pm$0.055, 3.250$\pm$0.043, 3.163$\pm$0.042, 3.173$\pm$0.055, 3.380$\pm$0.047, 3.515$\pm$0.056, 3.445$\pm$0.046 for TMC1, TMC2, TauRing, PMCMain, OMCA, OMCB and OMCL, respectively. Consequently, the mean and dispersion of the fitted values are $\rv^{\rm fit}=3.288\pm0.159$ for the seven sub-clouds (Table \ref{table:R_opt}).

On the other hand, if we define $\rbp \equiv A_{\Gbp}/E_{\Gbp,\ab{G}{RP}}$ and assume the extinction in a `very red' band, i.e. $W2$ band here is negligible, then,
\begin{equation}\label{equ:true_Rgbp}
\rbp \equiv \frac{A_{\Gbp}}{E_{\Gbp,\ab{G}{RP}}}\approx \frac{E_{\Gbp,W2}}{E_{\Gbp,\ab{G}{RP}}}
\end{equation}
the true $\rbp$ value can be estimated. It's 2.334$\pm$0.080, 2.312$\pm$0.060, 2.290$\pm$0.053, 2.320$\pm$0.065, 2.254$\pm$0.054, 2.285$\pm$0.056 and 2.319$\pm$0.054 respectively with the average $\rbp=2.302\pm0.027$. However, it should be noted that the results are the lower limit since the calculation omitted the extinction in W2. The linear fit result between $\rbp$ and $\rv^{\rm fit}$ is $\rbp =-2.836\rv^{\rm fit}+9.818$ with correlation coefficient of -0.486, indicating a weak anti-correlation.

This result is consistent with \citet{Li2023_TMCLaw}, which derives the average $\rv=3.13\pm0.32$ for the diffuse region in the Taurus MC with $E_{\Gbp,\ab{G}{RP}} <2$ mag. It is also similar to the results of \citet{Zhang2023_RvMap} except for OMCL, whose $\rv$ is around 3.8. In the $\rv$ map of \citet{Lee18_Rv}, these 7 substructures exhibit a similar $\rv$ range from 3.1 to 4.0.

\begin{figure*}
\centering
\includegraphics[width=\textwidth]{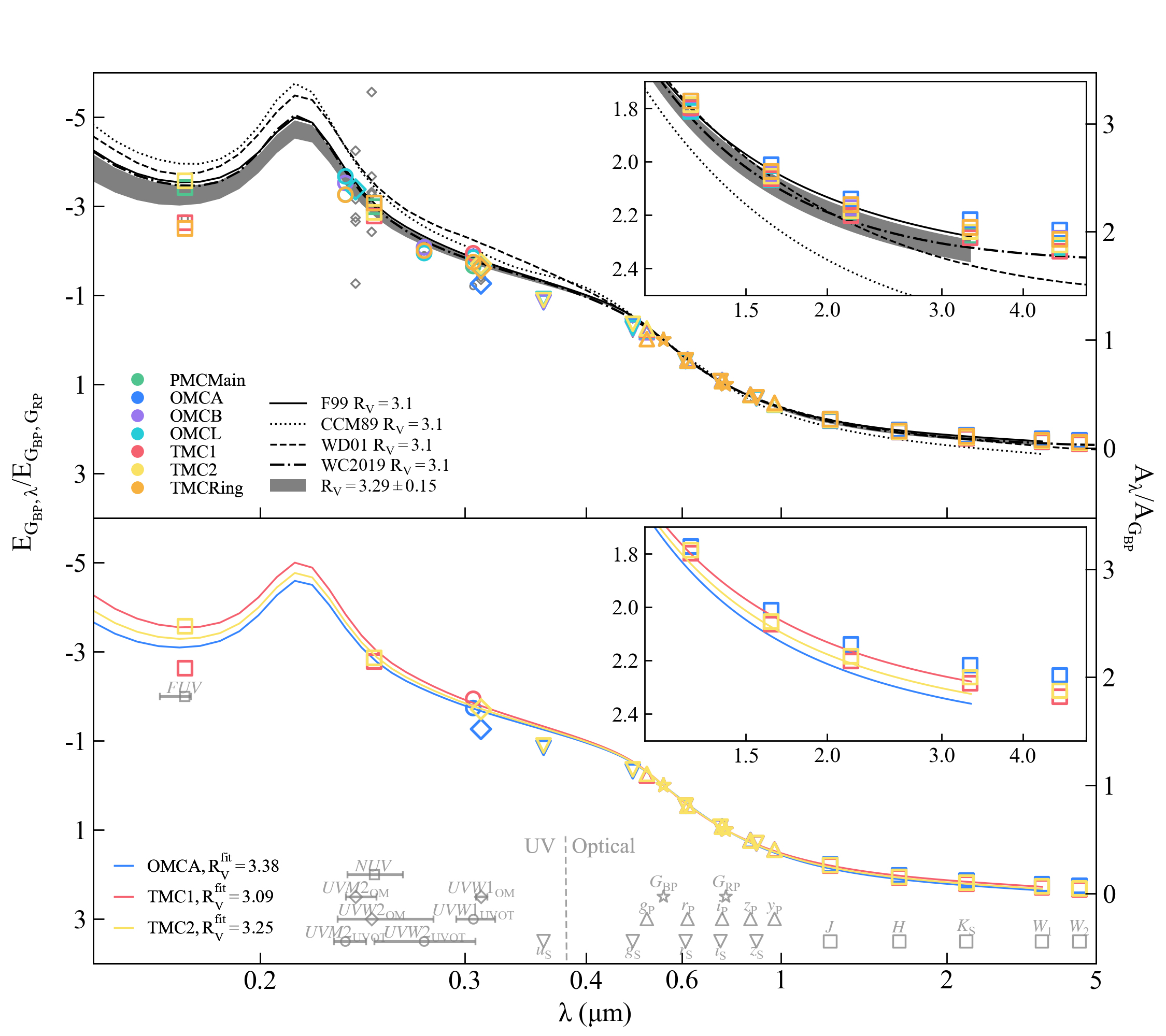}
\caption{The UV and optical extinction curves for the seven substructures. For comparison, the model $\rv=3.1$ extinction curves are added, including F99 \citep{F99}, WD01 \citep{WD01}, CCM89 \citep{CCM89} and WC2019 \citep{Wang19_law}. If the number of stars in the sample is fewer than 10, the marker becomes an open symbol. The left and right y-axis are the CER and converted $\ab{A}{\lambda}/\ab{A}{\Gbp}$ by Equation \ref{equ:AxAlambda}. The horizontal bar of the UV filters represent the range of effective wavelength consistent with that of stellar effective temperature.}
\label{fig:law_UVoptical}
\end{figure*}

\begin{figure*}
\centering
\includegraphics[width=\textwidth]{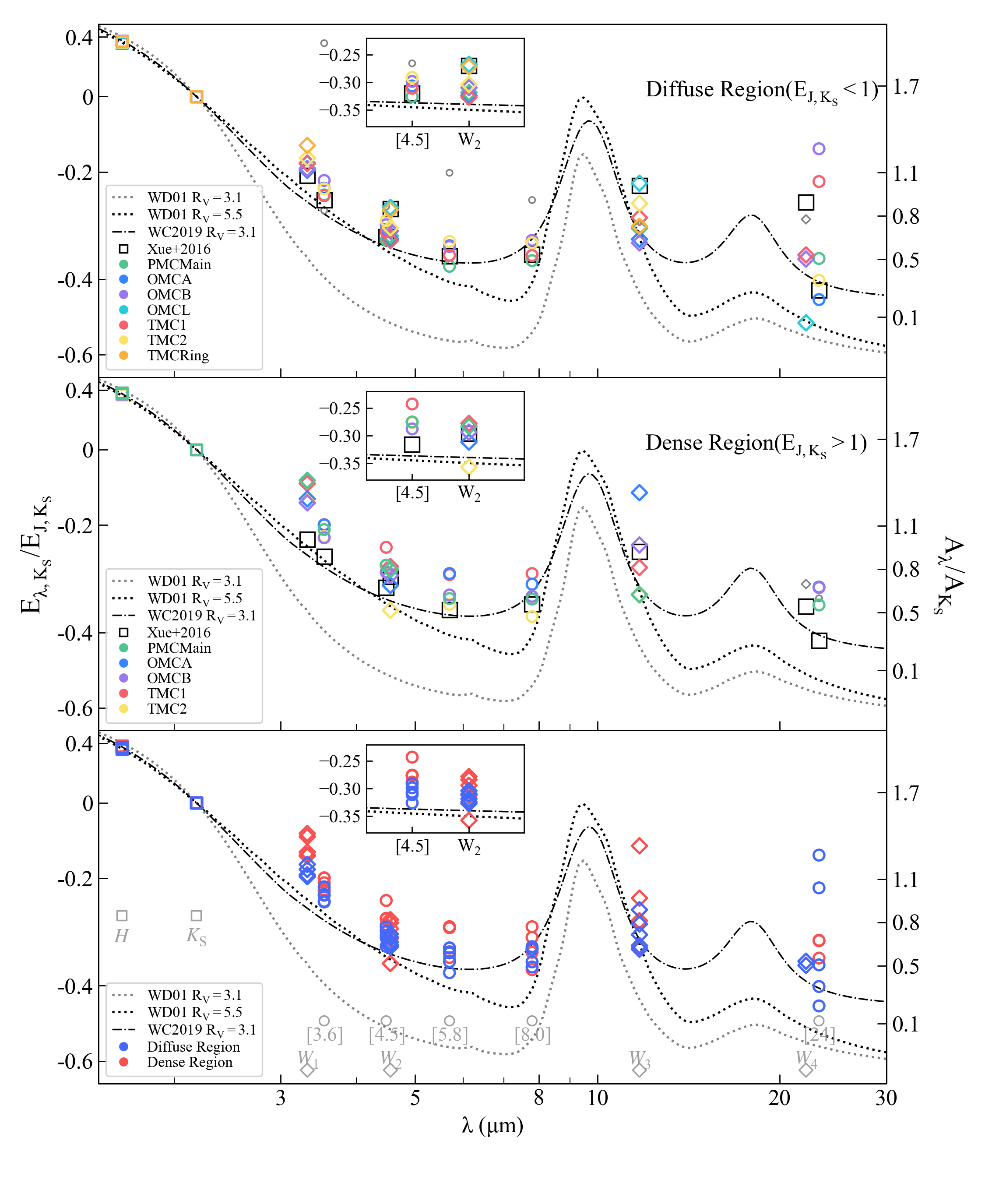}
\caption{The same as Figure \ref{fig:law_UVoptical}, but for the IR bands, where the color of the symbol denotes the sub-cloud, and the shape of the symbol denotes the sky survey. The top and middle panels show the extinction law in the region with $E_{J,\K}<1$ mag and $E_{J,\K}>1$ mag respectively, and the bottom panel compares the differences between the diffuse and dense regions. The comparison models include WC2019 \citep{Wang19_law}, WD01 \citep{WD01} with $\rv=3.1$ and $\rv=5.5$, and Xue+2016 \citep{Xue2016_IRlaw}.}
\label{fig:law_IR}
\end{figure*}

The extinction laws are shown in Figure \ref{fig:law_UVoptical}, where the top panel shows the CERs of 7 MCs, which coincide with the F99 $\rv=3.1$ and WC2019 curves, while it is below the WD01 and CCM89 curves. One outstanding waveband is FUV for TMC1 and TMCRing, which lies apparently below all the model curves. Specifically, as can be read from Table \ref{table:curve}, the value of $E(\Gbp-\mathrm{FUV})/E(\Gbp-\ab{G}{RP})$ is -3.4 and -3.5 for PMCMain and TMC2, respectively, which agree with the F99 curve, while -2.6 and -2.5 for TMC1 and TMCRing, respectively, apparently smaller (absolute value) than the F99 model. Even with the error taken into account, the difference exists. Interestingly, this value is -2.95 with the sample in the whole sky by \citet{Sun2021_UVmap}, just between the two groups. Since -2.95 represents the average over various environments, the bifurcation seems reasonable. However, it is difficult to understand that TMC1 and TMCRing are very different in environment but share similar value. The second panel of Figure \ref{fig:law_UVoptical} shows three extinction curves of TMC1, TMC2 and OMCA. From NIR band on, slight difference can be seen in the inset where the solid lines mark the fitted $\rv^{\rm fit}$ curve. For UV band, unfortunately, there is a lack of data for OMCA in the FUV and NUV band. Therefore, we compare TMC2 with TMC1, which have the largest difference in the FUV bands. It is found that they have small difference in $\rbp$ compared with the difference in the FUV band. This shows that the UV extinction exhibits significant variation and cannot be described by the $R$ value calculated from optical measurements.

\subsection{The IR Extinction Law}

As stated above, the IR extinction law is studied in diffuse and dense case separately with the APOGEE data as the maximum extinction reaches $E_{J,\K} \approx 4$, roughly equal to $\ab{A}{V}\approx24$ mag. Here, the borderline between `dense' and `diffuse' is set at $E_{J,\K}=1$ mag, i.e. $A_{\rm V} \approx 6$ mag. The results are displayed in Figure \ref{fig:law_IR}, where the top, middle and bottom panel is for the diffuse, dense, and diffuse plus dense region, respectively, and the inset intends to separate the $W2$ and [4.5] band clearly. The values from both diffuse and dense regions are more consistent with the WC2019 and WD01 $\rv=5.5$ model than $\rv=3.1$ one, which represents a `flat' mid-IR extinction law. In a quantitative analysis, residuals are calculated between our result and WC2019 and WD01 $\rv=5.5$, with the mean being -0.02 for WC2019 and -0.05 for WD01 $\rv=5.5$. \citet{Indebetouw2005_LawGLIMPSE} and \citet{Lutz96_LawGC} also show a flattening curve in mid-IR range. Additionally, there is substantial variation in the 9.7$\mu$m silicate absorption across these MCs, in that OMCL shows an outstanding absorption in silicate compared with other MCs.

The CERs for dense regions is systematically higher than that of \citet{Xue2016_IRlaw} that represents the average over various environments. \citet{wang24_jwstIR} describes the NIR extinction curve with a power-law $A_{\lambda}\propto \lambda^{-\alpha}$ with the average index $\alpha=1.98\pm 0.15$, which is slightly lower than our results. This may be caused by that their samples are from dense region in galactic center and mixed with multiple MCs. On the other hand, the result in the WISE/W4 and Spitzer/MIPS24 bands shows no such difference.

\subsection{Abnormal Extinction in the W1 band}

The W1-band extinction in the dense region is abnormally high as visible in the middle panel of Figure \ref{fig:law_IR} in comparison with the similar band [3.6] and the result of \citet{Xue2016_IRlaw}. By checking Table \ref{table:curve}, the value of $E_{W1,\K}/E_{J,\K}$ is around -0.1, while $E_{[3.6],\K}/E_{J,\K}$ around -0.2. This value is -0.24 in \citet{Xue2016_IRlaw}, consistent with our result at [3.6]. In addition, this feature is very strong in TMC1, TMC2 and PMCMain, and becomes weak in OMCA, OMCB.

\citet{Wang2013_coalsack} found that the $[4.5]$ extinction is abnormally high in Coalsack Nebula, and explained it by ice absorption. Besides, the regions with different water ice absorption in one MC are also found \citep{Boogert15_IcyUniverse, Whittet13_L183}. By comparing the filter transmission of W1 and [3.6] with the water ice absorption band around 3$\mu$m, it is found that the W1 band can be apparently influenced by the water ice while the [3.6] band is barely affected. Thus, the additional extinction in the W1 band may be attributed to the water ice absorption. The above analysis indicates that this absorption occurs only in the dense regions, i.e. $E_{J,\K}>1$ mag, which coincides with previous reports. In addition, the strength depends not only on the extinction since it is not so strong in the OMC cloud as in the TMC or PMC clouds. The radiation field may be the another factor. A further detailed study of this phenomenon is under way and will be presented in next paper.

\section{Summary}\label{sec:summary}
We measured the extinction law from UV to IR with multiband photometric data from XMM-OM, UVOT, GALEX, Gaia, SDSS, PS1, SDSS, 2MASS, SPITZER and WISE surveys, combined with spectral data from LAMOST and APOGEE surveys. The major results are as follows.

\begin{enumerate}
    \item The extinction law of diffuse regions toward  seven substructures in three nearby star-forming regions can generally be represented by $\rv^{\rm fit}=3.288\pm0.159$, and the true $\rbp$ values is $\rbp=2.302\pm0.027$.
    \item The UV extinction is compromised by the shallow depth (i.e. $\ab{A}{V} \leq 2$mag) of the measurements in this work, still it shows significant variation. One outstanding waveband is FUV. TMC1 and TMCRing are very different in environment but share similar $R$ value. The conventional $R$ parameter can not match the UV extinction law.
    \item In IR band, the extinction law generally follows the WC2019 model better than WD01 $R_{\rm V}=5.5$ model. The extinction law for dense regions is systematically higher compared with average results in \citet{Xue2016_IRlaw}.
    \item An abnormal high extinction is found in the dense region in WISE/W1 band. By comparing the filter transmission with water ice absorption around 3$\mu$m, the additional extinction in W1 band may be attributed to water ice absorption.
\end{enumerate}

\begin{acknowledgments}
We thank the referee for very helpful suggestions. This work is supported by the NSFC project 12133002 and 12373028, National Key R\&D Program of China No. 2019YFA0405500, and CMS-CSST-2021-A09. S.W. acknowledges support from the Youth Innovation Promotion Association of the CAS (grant No. 2023065). This work has made use of the data from LAMOST, APOGEE, XMM-OM, Swift/UVOT, GALEX, Gaia, SDSS, PS1, 2MASS, Spitzer, and WISE surveys.
\end{acknowledgments}	

\bibliographystyle{aasjournal}
\bibliography{law}

\begin{thebibliography}{}
\expandafter\ifx\csname natexlab\endcsname\relax\def\natexlab#1{#1}\fi
\providecommand{\url}[1]{\href{#1}{#1}}
\providecommand{\dodoi}[1]{doi:~\href{http://doi.org/#1}{\nolinkurl{#1}}}
\providecommand{\doeprint}[1]{\href{http://ascl.net/#1}{\nolinkurl{http://ascl.net/#1}}}
\providecommand{\doarXiv}[1]{\href{https://arxiv.org/abs/#1}{\nolinkurl{https://arxiv.org/abs/#1}}}

\bibitem[{{Abdurro'uf} {et~al.}(2022){Abdurro'uf}, {Accetta}, {Aerts}, {Silva
  Aguirre}, {Ahumada}, {Ajgaonkar}, {Filiz Ak}, {Alam}, {Allende Prieto},
  {Almeida}, {Anders}, {Anderson}, {Andrews}, {Anguiano}, {Aquino-Ort{\'\i}z},
  {Arag{\'o}n-Salamanca}, {Argudo-Fern{\'a}ndez}, {Ata}, {Aubert},
  {Avila-Reese}, {Badenes}, {Barb{\'a}}, {Barger}, {Barrera-Ballesteros},
  {Beaton}, {Beers}, {Belfiore}, {Bender}, {Bernardi}, {Bershady}, {Beutler},
  {Bidin}, {Bird}, {Bizyaev}, {Blanc}, {Blanton}, {Boardman}, {Bolton},
  {Boquien}, {Borissova}, {Bovy}, {Brandt}, {Brown}, {Brownstein}, {Brusa},
  {Buchner}, {Bundy}, {Burchett}, {Bureau}, {Burgasser}, {Cabang}, {Campbell},
  {Cappellari}, {Carlberg}, {Wanderley}, {Carrera}, {Cash}, {Chen}, {Chen},
  {Cherinka}, {Chiappini}, {Choi}, {Chojnowski}, {Chung}, {Clerc}, {Cohen},
  {Comerford}, {Comparat}, {da Costa}, {Covey}, {Crane}, {Cruz-Gonzalez},
  {Culhane}, {Cunha}, {Dai}, {Damke}, {Darling}, {Davidson}, {Davies},
  {Dawson}, {De Lee}, {Diamond-Stanic}, {Cano-D{\'\i}az}, {S{\'a}nchez},
  {Donor}, {Duckworth}, {Dwelly}, {Eisenstein}, {Elsworth}, {Emsellem},
  {Eracleous}, {Escoffier}, {Fan}, {Farr}, {Feng}, {Fern{\'a}ndez-Trincado},
  {Feuillet}, {Filipp}, {Fillingham}, {Frinchaboy}, {Fromenteau}, {Galbany},
  {Garc{\'\i}a}, {Garc{\'\i}a-Hern{\'a}ndez}, {Ge}, {Geisler}, {Gelfand},
  {G{\'e}ron}, {Gibson}, {Goddy}, {Godoy-Rivera}, {Grabowski}, {Green},
  {Greener}, {Grier}, {Griffith}, {Guo}, {Guy}, {Hadjara}, {Harding},
  {Hasselquist}, {Hayes}, {Hearty}, {Hern{\'a}ndez}, {Hill}, {Hogg},
  {Holtzman}, {Horta}, {Hsieh}, {Hsu}, {Hsu}, {Huber}, {Huertas-Company},
  {Hutchinson}, {Hwang}, {Ibarra-Medel}, {Chitham}, {Ilha}, {Imig}, {Jaekle},
  {Jayasinghe}, {Ji}, {Johnson}, {Jones}, {J{\"o}nsson}, {Katkov}, {Khalatyan},
  {Kinemuchi}, {Kisku}, {Knapen}, {Kneib}, {Kollmeier}, {Kong}, {Kounkel},
  {Kreckel}, {Krishnarao}, {Lacerna}, {Lane}, {Langgin}, {Lavender}, {Law},
  {Lazarz}, {Leung}, {Leung}, {Lewis}, {Li}, {Li}, {Lian}, {Liang}, {Lin},
  {Lin}, {Lin}, {Lintott}, {Long}, {Longa-Pe{\~n}a}, {L{\'o}pez-Cob{\'a}},
  {Lu}, {Lundgren}, {Luo}, {Mackereth}, {de la Macorra}, {Mahadevan},
  {Majewski}, {Manchado}, {Mandeville}, {Maraston}, {Margalef-Bentabol},
  {Masseron}, {Masters}, {Mathur}, {McDermid}, {Mckay}, {Merloni},
  {Merrifield}, {Meszaros}, {Miglio}, {Di Mille}, {Minniti}, {Minsley},
  {Monachesi}, {Moon}, {Mosser}, {Mulchaey}, {Muna}, {Mu{\~n}oz}, {Myers},
  {Myers}, {Nadathur}, {Nair}, {Nandra}, {Neumann}, {Newman}, {Nidever},
  {Nikakhtar}, {Nitschelm}, {O'Connell}, {Garma-Oehmichen}, {Luan Souza de
  Oliveira}, {Olney}, {Oravetz}, {Ortigoza-Urdaneta}, {Osorio}, {Otter},
  {Pace}, {Padilla}, {Pan}, {Pan}, {Parikh}, {Parker}, {Peirani}, {Pe{\~n}a
  Ram{\'\i}rez}, {Penny}, {Percival}, {Perez-Fournon}, {Pinsonneault},
  {Poidevin}, {Poovelil}, {Price-Whelan}, {B{\'a}rbara de Andrade Queiroz},
  {Raddick}, {Ray}, {Rembold}, {Riddle}, {Riffel}, {Riffel}, {Rix}, {Robin},
  {Rodr{\'\i}guez-Puebla}, {Roman-Lopes}, {Rom{\'a}n-Z{\'u}{\~n}iga}, {Rose},
  {Ross}, {Rossi}, {Rubin}, {Salvato}, {S{\'a}nchez}, {S{\'a}nchez-Gallego},
  {Sanderson}, {Santana Rojas}, {Sarceno}, {Sarmiento}, {Sayres}, {Sazonova},
  {Schaefer}, {Schiavon}, {Schlegel}, {Schneider}, {Schultheis}, {Schwope},
  {Serenelli}, {Serna}, {Shao}, {Shapiro}, {Sharma}, {Shen}, {Shetrone}, {Shu},
  {Simon}, {Skrutskie}, {Smethurst}, {Smith}, {Sobeck}, {Spoo}, {Sprague},
  {Stark}, {Stassun}, {Steinmetz}, {Stello}, {Stone-Martinez},
  {Storchi-Bergmann}, {Stringfellow}, {Stutz}, {Su}, {Taghizadeh-Popp},
  {Talbot}, {Tayar}, {Telles}, {Teske}, {Thakar}, {Theissen}, {Tkachenko},
  {Thomas}, {Tojeiro}, {Hernandez Toledo}, {Troup}, {Trump}, {Trussler},
  {Turner}, {Tuttle}, {Unda-Sanzana}, {V{\'a}zquez-Mata}, {Valentini},
  {Valenzuela}, {Vargas-Gonz{\'a}lez}, {Vargas-Maga{\~n}a}, {Alfaro},
  {Villanova}, {Vincenzo}, {Wake}, {Warfield}, {Washington}, {Weaver},
  {Weijmans}, {Weinberg}, {Weiss}, {Westfall}, {Wild}, {Wilde}, {Wilson},
  {Wilson}, {Wilson}, {Wolf}, {Wood-Vasey}, {Yan}, {Zamora}, {Zasowski},
  {Zhang}, {Zhao}, {Zheng}, {Zheng}, \& {Zhu}}]{APGDR17}
{Abdurro'uf}, {Accetta}, K., {Aerts}, C., {et~al.} 2022, \apjs, 259, 35,
  \dodoi{10.3847/1538-4365/ac4414}

\bibitem[{{Ahumada} {et~al.}(2020){Ahumada}, {Allende Prieto}, {Almeida},
  {Anders}, {Anderson}, {Andrews}, {Anguiano}, {Arcodia}, {Armengaud},
  {Aubert}, {Avila}, {Avila-Reese}, {Badenes}, {Balland}, {Barger},
  {Barrera-Ballesteros}, {Basu}, {Bautista}, {Beaton}, {Beers}, {Benavides},
  {Bender}, {Bernardi}, {Bershady}, {Beutler}, {Bidin}, {Bird}, {Bizyaev},
  {Blanc}, {Blanton}, {Boquien}, {Borissova}, {Bovy}, {Brandt}, {Brinkmann},
  {Brownstein}, {Bundy}, {Bureau}, {Burgasser}, {Burtin}, {Cano-D{\'\i}az},
  {Capasso}, {Cappellari}, {Carrera}, {Chabanier}, {Chaplin}, {Chapman},
  {Cherinka}, {Chiappini}, {Doohyun Choi}, {Chojnowski}, {Chung}, {Clerc},
  {Coffey}, {Comerford}, {Comparat}, {da Costa}, {Cousinou}, {Covey}, {Crane},
  {Cunha}, {Ilha}, {Dai}, {Damsted}, {Darling}, {Davidson}, {Davies}, {Dawson},
  {De}, {de la Macorra}, {De Lee}, {Queiroz}, {Deconto Machado}, {de la Torre},
  {Dell'Agli}, {du Mas des Bourboux}, {Diamond-Stanic}, {Dillon}, {Donor},
  {Drory}, {Duckworth}, {Dwelly}, {Ebelke}, {Eftekharzadeh}, {Davis Eigenbrot},
  {Elsworth}, {Eracleous}, {Erfanianfar}, {Escoffier}, {Fan}, {Farr},
  {Fern{\'a}ndez-Trincado}, {Feuillet}, {Finoguenov}, {Fofie},
  {Fraser-McKelvie}, {Frinchaboy}, {Fromenteau}, {Fu}, {Galbany}, {Garcia},
  {Garc{\'\i}a-Hern{\'a}ndez}, {Garma Oehmichen}, {Ge}, {Geimba Maia},
  {Geisler}, {Gelfand}, {Goddy}, {Gonzalez-Perez}, {Grabowski}, {Green},
  {Grier}, {Guo}, {Guy}, {Harding}, {Hasselquist}, {Hawken}, {Hayes}, {Hearty},
  {Hekker}, {Hogg}, {Holtzman}, {Horta}, {Hou}, {Hsieh}, {Huber}, {Hunt}, {Ider
  Chitham}, {Imig}, {Jaber}, {Jimenez Angel}, {Johnson}, {Jones},
  {J{\"o}nsson}, {Jullo}, {Kim}, {Kinemuchi}, {Kirkpatrick}, {Kite}, {Klaene},
  {Kneib}, {Kollmeier}, {Kong}, {Kounkel}, {Krishnarao}, {Lacerna}, {Lan},
  {Lane}, {Law}, {Le Goff}, {Leung}, {Lewis}, {Li}, {Lian}, {Lin}, {Long},
  {Longa-Pe{\~n}a}, {Lundgren}, {Lyke}, {Mackereth}, {MacLeod}, {Majewski},
  {Manchado}, {Maraston}, {Martini}, {Masseron}, {Masters}, {Mathur},
  {McDermid}, {Merloni}, {Merrifield}, {M{\'e}sz{\'a}ros}, {Miglio}, {Minniti},
  {Minsley}, {Miyaji}, {Mohammad}, {Mosser}, {Mueller}, {Muna},
  {Mu{\~n}oz-Guti{\'e}rrez}, {Myers}, {Nadathur}, {Nair}, {Nandra}, {Correa do
  Nascimento}, {Nevin}, {Newman}, {Nidever}, {Nitschelm}, {Noterdaeme},
  {O'Connell}, {Olmstead}, {Oravetz}, {Oravetz}, {Osorio}, {Pace}, {Padilla},
  {Palanque-Delabrouille}, {Palicio}, {Pan}, {Pan}, {Parker}, {Paviot},
  {Peirani}, {Ram{\'r}ez}, {Penny}, {Percival}, {Perez-Fournon},
  {P{\'e}rez-R{\`a}fols}, {Petitjean}, {Pieri}, {Pinsonneault}, {Poovelil},
  {Povick}, {Prakash}, {Price-Whelan}, {Raddick}, {Raichoor}, {Ray}, {Rembold},
  {Rezaie}, {Riffel}, {Riffel}, {Rix}, {Robin}, {Roman-Lopes},
  {Rom{\'a}n-Z{\'u}{\~n}iga}, {Rose}, {Ross}, {Rossi}, {Rowlands}, {Rubin},
  {Salvato}, {S{\'a}nchez}, {S{\'a}nchez-Menguiano}, {S{\'a}nchez-Gallego},
  {Sayres}, {Schaefer}, {Schiavon}, {Schimoia}, {Schlafly}, {Schlegel},
  {Schneider}, {Schultheis}, {Schwope}, {Seo}, {Serenelli}, {Shafieloo},
  {Shamsi}, {Shao}, {Shen}, {Shetrone}, {Shirley}, {Silva Aguirre}, {Simon},
  {Skrutskie}, {Slosar}, {Smethurst}, {Sobeck}, {Sodi}, {Souto}, {Stark},
  {Stassun}, {Steinmetz}, {Stello}, {Stermer}, {Storchi-Bergmann},
  {Streblyanska}, {Stringfellow}, {Stutz}, {Su{\'a}rez}, {Sun},
  {Taghizadeh-Popp}, {Talbot}, {Tayar}, {Thakar}, {Theriault}, {Thomas},
  {Thomas}, {Tinker}, {Tojeiro}, {Toledo}, {Tremonti}, {Troup}, {Tuttle},
  {Unda-Sanzana}, {Valentini}, {Vargas-Gonz{\'a}lez}, {Vargas-Maga{\~n}a},
  {V{\'a}zquez-Mata}, {Vivek}, {Wake}, {Wang}, {Weaver}, {Weijmans}, {Wild},
  {Wilson}, {Wilson}, {Wolthuis}, {Wood-Vasey}, {Yan}, {Yang}, {Y{\`e}che},
  {Zamora}, {Zarrouk}, {Zasowski}, {Zhang}, {Zhao}, {Zhao}, {Zheng}, {Zheng},
  {Zhu}, \& {Zou}}]{SDSS16}
{Ahumada}, R., {Allende Prieto}, C., {Almeida}, A., {et~al.} 2020, \apjs, 249,
  3, \dodoi{10.3847/1538-4365/ab929e}

\bibitem[{{Bailer-Jones} {et~al.}(2021){Bailer-Jones}, {Rybizki}, {Fouesneau},
  {Demleitner}, \& {Andrae}}]{BJdistance}
{Bailer-Jones}, C.~A.~L., {Rybizki}, J., {Fouesneau}, M., {Demleitner}, M., \&
  {Andrae}, R. 2021, \aj, 161, 147, \dodoi{10.3847/1538-3881/abd806}

\bibitem[{{Bianchi} {et~al.}(2017){Bianchi}, {Shiao}, \& {Thilker}}]{GALEX}
{Bianchi}, L., {Shiao}, B., \& {Thilker}, D. 2017, \apjs, 230, 24,
  \dodoi{10.3847/1538-4365/aa7053}

\bibitem[{{Boogert} {et~al.}(2015){Boogert}, {Gerakines}, \&
  {Whittet}}]{Boogert15_IcyUniverse}
{Boogert}, A.~C.~A., {Gerakines}, P.~A., \& {Whittet}, D. C.~B. 2015, \araa,
  53, 541, \dodoi{10.1146/annurev-astro-082214-122348}

\bibitem[{{Cao} {et~al.}(2023){Cao}, {Jiang}, {Zhao}, \& {Sun}}]{Cao23}
{Cao}, Z., {Jiang}, B., {Zhao}, H., \& {Sun}, M. 2023, \apj, 945, 132,
  \dodoi{10.3847/1538-4357/acbbc7}

\bibitem[{{Cardelli} {et~al.}(1989){Cardelli}, {Clayton}, \& {Mathis}}]{CCM89}
{Cardelli}, J.~A., {Clayton}, G.~C., \& {Mathis}, J.~S. 1989, \apj, 345, 245,
  \dodoi{10.1086/167900}

\bibitem[{{Carey} {et~al.}(2009){Carey}, {Noriega-Crespo}, {Mizuno}, {Shenoy},
  {Paladini}, {Kraemer}, {Price}, {Flagey}, {Ryan}, {Ingalls}, {Kuchar},
  {Pinheiro Gon{\c{c}}alves}, {Indebetouw}, {Billot}, {Marleau}, {Padgett},
  {Rebull}, {Bressert}, {Ali}, {Molinari}, {Martin}, {Berriman}, {Boulanger},
  {Latter}, {Miville-Deschenes}, {Shipman}, \& {Testi}}]{MIPSGAL}
{Carey}, S.~J., {Noriega-Crespo}, A., {Mizuno}, D.~R., {et~al.} 2009, \pasp,
  121, 76, \dodoi{10.1086/596581}

\bibitem[{{Chen} {et~al.}(2019){Chen}, {Huang}, {Yuan}, {Wang}, {Fan}, {Xiang},
  {Zhang}, {Tian}, \& {Liu}}]{Chen19_dustmap}
{Chen}, B.~Q., {Huang}, Y., {Yuan}, H.~B., {et~al.} 2019, \mnras, 483, 4277,
  \dodoi{10.1093/mnras/sty3341}

\bibitem[{{Churchwell} {et~al.}(2009){Churchwell}, {Babler}, {Meade},
  {Whitney}, {Benjamin}, {Indebetouw}, {Cyganowski}, {Robitaille}, {Povich},
  {Watson}, \& {Bracker}}]{GLIMPSE}
{Churchwell}, E., {Babler}, B.~L., {Meade}, M.~R., {et~al.} 2009, \pasp, 121,
  213, \dodoi{10.1086/597811}

\bibitem[{{Cutri} {et~al.}(2021){Cutri}, {Wright}, {Conrow}, {Fowler},
  {Eisenhardt}, {Grillmair}, {Kirkpatrick}, {Masci}, {McCallon}, {Wheelock},
  {Fajardo-Acosta}, {Yan}, {Benford}, {Harbut}, {Jarrett}, {Lake}, {Leisawitz},
  {Ressler}, {Stanford}, {Tsai}, {Liu}, {Helou}, {Mainzer}, {Gettngs},
  {Gonzalez}, {Hoffman}, {Marsh}, {Padgett}, {Skrutskie}, {Beck}, {Papin}, \&
  {Wittman}}]{WISE}
{Cutri}, R.~M., {Wright}, E.~L., {Conrow}, T., {et~al.} 2021, VizieR Online
  Data Catalog, II/328

\bibitem[{{Ducati} {et~al.}(2001){Ducati}, {Bevilacqua}, {Rembold}, \&
  {Ribeiro}}]{Ducati_2001_BE}
{Ducati}, J.~R., {Bevilacqua}, C.~M., {Rembold}, S.~B., \& {Ribeiro}, D. 2001,
  \apj, 558, 309, \dodoi{10.1086/322439}

\bibitem[{{Evans} {et~al.}(2003){Evans}, {Allen}, {Blake}, {Boogert}, {Bourke},
  {Harvey}, {Kessler}, {Koerner}, {Lee}, {Mundy}, {Myers}, {Padgett},
  {Pontoppidan}, {Sargent}, {Stapelfeldt}, {van Dishoeck}, {Young}, \&
  {Young}}]{c2d}
{Evans}, Neal~J., I., {Allen}, L.~E., {Blake}, G.~A., {et~al.} 2003, \pasp,
  115, 965, \dodoi{10.1086/376697}

\bibitem[{{Evans} {et~al.}(2009){Evans}, {Dunham}, {J{\o}rgensen}, {Enoch},
  {Mer{\'\i}n}, {van Dishoeck}, {Alcal{\'a}}, {Myers}, {Stapelfeldt}, {Huard},
  {Allen}, {Harvey}, {van Kempen}, {Blake}, {Koerner}, {Mundy}, {Padgett}, \&
  {Sargent}}]{c2d_SFR}
{Evans}, Neal~J., I., {Dunham}, M.~M., {J{\o}rgensen}, J.~K., {et~al.} 2009,
  \apjs, 181, 321, \dodoi{10.1088/0067-0049/181/2/321}

\bibitem[{{Fazio} {et~al.}(2004){Fazio}, {Hora}, {Allen}, {Ashby}, {Barmby},
  {Deutsch}, {Huang}, {Kleiner}, {Marengo}, {Megeath}, {Melnick}, {Pahre},
  {Patten}, {Polizotti}, {Smith}, {Taylor}, {Wang}, {Willner}, {Hoffmann},
  {Pipher}, {Forrest}, {McMurty}, {McCreight}, {McKelvey}, {McMurray}, {Koch},
  {Moseley}, {Arendt}, {Mentzell}, {Marx}, {Losch}, {Mayman}, {Eichhorn},
  {Krebs}, {Jhabvala}, {Gezari}, {Fixsen}, {Flores}, {Shakoorzadeh}, {Jungo},
  {Hakun}, {Workman}, {Karpati}, {Kichak}, {Whitley}, {Mann}, {Tollestrup},
  {Eisenhardt}, {Stern}, {Gorjian}, {Bhattacharya}, {Carey}, {Nelson},
  {Glaccum}, {Lacy}, {Lowrance}, {Laine}, {Reach}, {Stauffer}, {Surace},
  {Wilson}, {Wright}, {Hoffman}, {Domingo}, \& {Cohen}}]{IRAC}
{Fazio}, G.~G., {Hora}, J.~L., {Allen}, L.~E., {et~al.} 2004, \apjs, 154, 10,
  \dodoi{10.1086/422843}

\bibitem[{{Fitzpatrick}(1999)}]{F99}
{Fitzpatrick}, E.~L. 1999, \pasp, 111, 63, \dodoi{10.1086/316293}

\bibitem[{{Foster} {et~al.}(2013){Foster}, {Mandel}, {Pineda}, {Covey}, {Arce},
  \& {Goodman}}]{Foster2013_PMClaw}
{Foster}, J.~B., {Mandel}, K.~S., {Pineda}, J.~E., {et~al.} 2013, \mnras, 428,
  1606, \dodoi{10.1093/mnras/sts144}

\bibitem[{{Gaia Collaboration} {et~al.}(2023){Gaia Collaboration}, {Vallenari},
  {Brown}, {Prusti}, {de Bruijne}, {Arenou}, {Babusiaux}, {Biermann},
  {Creevey}, {Ducourant}, {Evans}, {Eyer}, {Guerra}, {Hutton}, {Jordi},
  {Klioner}, {Lammers}, {Lindegren}, {Luri}, {Mignard}, {Panem}, {Pourbaix},
  {Randich}, {Sartoretti}, {Soubiran}, {Tanga}, {Walton}, {Bailer-Jones},
  {Bastian}, {Drimmel}, {Jansen}, {Katz}, {Lattanzi}, {van Leeuwen}, {Bakker},
  {Cacciari}, {Casta{\~n}eda}, {De Angeli}, {Fabricius}, {Fouesneau},
  {Fr{\'e}mat}, {Galluccio}, {Guerrier}, {Heiter}, {Masana}, {Messineo},
  {Mowlavi}, {Nicolas}, {Nienartowicz}, {Pailler}, {Panuzzo}, {Riclet}, {Roux},
  {Seabroke}, {Sordo}, {Th{\'e}venin}, {Gracia-Abril}, {Portell}, {Teyssier},
  {Altmann}, {Andrae}, {Audard}, {Bellas-Velidis}, {Benson}, {Berthier},
  {Blomme}, {Burgess}, {Busonero}, {Busso}, {C{\'a}novas}, {Carry}, {Cellino},
  {Cheek}, {Clementini}, {Damerdji}, {Davidson}, {de Teodoro}, {Nu{\~n}ez
  Campos}, {Delchambre}, {Dell'Oro}, {Esquej}, {Fern{\'a}ndez-Hern{\'a}ndez},
  {Fraile}, {Garabato}, {Garc{\'\i}a-Lario}, {Gosset}, {Haigron}, {Halbwachs},
  {Hambly}, {Harrison}, {Hern{\'a}ndez}, {Hestroffer}, {Hodgkin}, {Holl},
  {Jan{\ss}en}, {Jevardat de Fombelle}, {Jordan}, {Krone-Martins}, {Lanzafame},
  {L{\"o}ffler}, {Marchal}, {Marrese}, {Moitinho}, {Muinonen}, {Osborne},
  {Pancino}, {Pauwels}, {Recio-Blanco}, {Reyl{\'e}}, {Riello}, {Rimoldini},
  {Roegiers}, {Rybizki}, {Sarro}, {Siopis}, {Smith}, {Sozzetti}, {Utrilla},
  {van Leeuwen}, {Abbas}, {{\'A}brah{\'a}m}, {Abreu Aramburu}, {Aerts},
  {Aguado}, {Ajaj}, {Aldea-Montero}, {Altavilla}, {{\'A}lvarez}, {Alves},
  {Anders}, {Anderson}, {Anglada Varela}, {Antoja}, {Baines}, {Baker},
  {Balaguer-N{\'u}{\~n}ez}, {Balbinot}, {Balog}, {Barache}, {Barbato},
  {Barros}, {Barstow}, {Bartolom{\'e}}, {Bassilana}, {Bauchet}, {Becciani},
  {Bellazzini}, {Berihuete}, {Bernet}, {Bertone}, {Bianchi}, {Binnenfeld},
  {Blanco-Cuaresma}, {Blazere}, {Boch}, {Bombrun}, {Bossini}, {Bouquillon},
  {Bragaglia}, {Bramante}, {Breedt}, {Bressan}, {Brouillet}, {Brugaletta},
  {Bucciarelli}, {Burlacu}, {Butkevich}, {Buzzi}, {Caffau}, {Cancelliere},
  {Cantat-Gaudin}, {Carballo}, {Carlucci}, {Carnerero}, {Carrasco},
  {Casamiquela}, {Castellani}, {Castro-Ginard}, {Chaoul}, {Charlot}, {Chemin},
  {Chiaramida}, {Chiavassa}, {Chornay}, {Comoretto}, {Contursi}, {Cooper},
  {Cornez}, {Cowell}, {Crifo}, {Cropper}, {Crosta}, {Crowley}, {Dafonte},
  {Dapergolas}, {David}, {David}, {de Laverny}, {De Luise}, {De March}, {De
  Ridder}, {de Souza}, {de Torres}, {del Peloso}, {del Pozo}, {Delbo},
  {Delgado}, {Delisle}, {Demouchy}, {Dharmawardena}, {Di Matteo}, {Diakite},
  {Diener}, {Distefano}, {Dolding}, {Edvardsson}, {Enke}, {Fabre}, {Fabrizio},
  {Faigler}, {Fedorets}, {Fernique}, {Fienga}, {Figueras}, {Fournier},
  {Fouron}, {Fragkoudi}, {Gai}, {Garcia-Gutierrez}, {Garcia-Reinaldos},
  {Garc{\'\i}a-Torres}, {Garofalo}, {Gavel}, {Gavras}, {Gerlach}, {Geyer},
  {Giacobbe}, {Gilmore}, {Girona}, {Giuffrida}, {Gomel}, {Gomez},
  {Gonz{\'a}lez-N{\'u}{\~n}ez}, {Gonz{\'a}lez-Santamar{\'\i}a},
  {Gonz{\'a}lez-Vidal}, {Granvik}, {Guillout}, {Guiraud},
  {Guti{\'e}rrez-S{\'a}nchez}, {Guy}, {Hatzidimitriou}, {Hauser}, {Haywood},
  {Helmer}, {Helmi}, {Sarmiento}, {Hidalgo}, {Hilger}, {H{\l}adczuk}, {Hobbs},
  {Holland}, {Huckle}, {Jardine}, {Jasniewicz}, {Jean-Antoine Piccolo},
  {Jim{\'e}nez-Arranz}, {Jorissen}, {Juaristi Campillo}, {Julbe}, {Karbevska},
  {Kervella}, {Khanna}, {Kontizas}, {Kordopatis}, {Korn}, {K{\'o}sp{\'a}l},
  {Kostrzewa-Rutkowska}, {Kruszy{\'n}ska}, {Kun}, {Laizeau}, {Lambert},
  {Lanza}, {Lasne}, {Le Campion}, {Lebreton}, {Lebzelter}, {Leccia}, {Leclerc},
  {Lecoeur-Taibi}, {Liao}, {Licata}, {Lindstr{\o}m}, {Lister}, {Livanou},
  {Lobel}, {Lorca}, {Loup}, {Madrero Pardo}, {Magdaleno Romeo}, {Managau},
  {Mann}, {Manteiga}, {Marchant}, {Marconi}, {Marcos}, {Marcos Santos},
  {Mar{\'\i}n Pina}, {Marinoni}, {Marocco}, {Marshall}, {Martin Polo},
  {Mart{\'\i}n-Fleitas}, {Marton}, {Mary}, {Masip}, {Massari},
  {Mastrobuono-Battisti}, {Mazeh}, {McMillan}, {Messina}, {Michalik}, {Millar},
  {Mints}, {Molina}, {Molinaro}, {Moln{\'a}r}, {Monari}, {Mongui{\'o}},
  {Montegriffo}, {Montero}, {Mor}, {Mora}, {Morbidelli}, {Morel}, {Morris},
  {Muraveva}, {Murphy}, {Musella}, {Nagy}, {Noval}, {Oca{\~n}a}, {Ogden},
  {Ordenovic}, {Osinde}, {Pagani}, {Pagano}, {Palaversa}, {Palicio},
  {Pallas-Quintela}, {Panahi}, {Payne-Wardenaar}, {Pe{\~n}alosa Esteller},
  {Penttil{\"a}}, {Pichon}, {Piersimoni}, {Pineau}, {Plachy}, {Plum}, {Poggio},
  {Pr{\v{s}}a}, {Pulone}, {Racero}, {Ragaini}, {Rainer}, {Raiteri}, {Rambaux},
  {Ramos}, {Ramos-Lerate}, {Re Fiorentin}, {Regibo}, {Richards}, {Rios Diaz},
  {Ripepi}, {Riva}, {Rix}, {Rixon}, {Robichon}, {Robin}, {Robin}, {Roelens},
  {Rogues}, {Rohrbasser}, {Romero-G{\'o}mez}, {Rowell}, {Royer}, {Ruz Mieres},
  {Rybicki}, {Sadowski}, {S{\'a}ez N{\'u}{\~n}ez}, {Sagrist{\`a} Sell{\'e}s},
  {Sahlmann}, {Salguero}, {Samaras}, {Sanchez Gimenez}, {Sanna},
  {Santove{\~n}a}, {Sarasso}, {Schultheis}, {Sciacca}, {Segol}, {Segovia},
  {S{\'e}gransan}, {Semeux}, {Shahaf}, {Siddiqui}, {Siebert}, {Siltala},
  {Silvelo}, {Slezak}, {Slezak}, {Smart}, {Snaith}, {Solano}, {Solitro},
  {Souami}, {Souchay}, {Spagna}, {Spina}, {Spoto}, {Steele},
  {Steidelm{\"u}ller}, {Stephenson}, {S{\"u}veges}, {Surdej}, {Szabados},
  {Szegedi-Elek}, {Taris}, {Taylor}, {Teixeira}, {Tolomei}, {Tonello}, {Torra},
  {Torra}, {Torralba Elipe}, {Trabucchi}, {Tsounis}, {Turon}, {Ulla}, {Unger},
  {Vaillant}, {van Dillen}, {van Reeven}, {Vanel}, {Vecchiato}, {Viala},
  {Vicente}, {Voutsinas}, {Weiler}, {Wevers}, {Wyrzykowski}, {Yoldas}, {Yvard},
  {Zhao}, {Zorec}, {Zucker}, \& {Zwitter}}]{GaiaDR3}
{Gaia Collaboration}, {Vallenari}, A., {Brown}, A.~G.~A., {et~al.} 2023, \aap,
  674, A1, \dodoi{10.1051/0004-6361/202243940}

\bibitem[{{Hodapp} {et~al.}(2004){Hodapp}, {Kaiser}, {Aussel}, {Burgett},
  {Chambers}, {Chun}, {Dombeck}, {Douglas}, {Hafner}, {Heasley}, {Hoblitt},
  {Hude}, {Isani}, {Jedicke}, {Jewitt}, {Laux}, {Luppino}, {Lupton}, {Maberry},
  {Magnier}, {Mannery}, {Monet}, {Morgan}, {Onaka}, {Price}, {Ryan},
  {Siegmund}, {Szapudi}, {Tonry}, {Wainscoat}, \& {Waterson}}]{PS1}
{Hodapp}, K.~W., {Kaiser}, N., {Aussel}, H., {et~al.} 2004, Astronomische
  Nachrichten, 325, 636, \dodoi{10.1002/asna.200410300}

\bibitem[{{Indebetouw} {et~al.}(2005){Indebetouw}, {Mathis}, {Babler}, {Meade},
  {Watson}, {Whitney}, {Wolff}, {Wolfire}, {Cohen}, {Bania}, {Benjamin},
  {Clemens}, {Dickey}, {Jackson}, {Kobulnicky}, {Marston}, {Mercer},
  {Stauffer}, {Stolovy}, \& {Churchwell}}]{Indebetouw2005_LawGLIMPSE}
{Indebetouw}, R., {Mathis}, J.~S., {Babler}, B.~L., {et~al.} 2005, \apj, 619,
  931, \dodoi{10.1086/426679}

\bibitem[{{Jian} {et~al.}(2017){Jian}, {Gao}, {Zhao}, \&
  {Jiang}}]{Jian2017_RevBE}
{Jian}, M., {Gao}, S., {Zhao}, H., \& {Jiang}, B. 2017, \aj, 153, 5,
  \dodoi{10.3847/1538-3881/153/1/5}

\bibitem[{{Lee} {et~al.}(2018){Lee}, {Green}, {Schlafly}, {Finkbeiner},
  {Burgett}, {Chambers}, {Flewelling}, {Hodapp}, {Kaiser}, {Kudritzki},
  {Magnier}, {Metcalfe}, {Wainscoat}, \& {Waters}}]{Lee18_Rv}
{Lee}, A., {Green}, G.~M., {Schlafly}, E.~F., {et~al.} 2018, \apj, 854, 79,
  \dodoi{10.3847/1538-4357/aaaa6d}

\bibitem[{{Li} {et~al.}(2024){Li}, {Chen}, {Jiang}, {Gao}, \&
  {Chen}}]{Li2024_DustGrowth}
{Li}, J., {Chen}, B., {Jiang}, B., {Gao}, J., \& {Chen}, X. 2024, \apjl, 968,
  L26, \dodoi{10.3847/2041-8213/ad54c7}

\bibitem[{{Li} {et~al.}(2023){Li}, {Wang}, {Chen}, \& {Jiang}}]{Li2023_TMCLaw}
{Li}, L., {Wang}, S., {Chen}, X., \& {Jiang}, Q. 2023, \apj, 956, 26,
  \dodoi{10.3847/1538-4357/aced8a}

\bibitem[{{Luo} {et~al.}(2015){Luo}, {Zhao}, {Zhao}, {Deng}, {Liu}, {Jing},
  {Wang}, {Zhang}, {Shi}, {Cui}, {Chu}, {Li}, {Bai}, {Wu}, {Cai}, {Cao}, {Cao},
  {Carlin}, {Chen}, {Chen}, {Chen}, {Chen}, {Chen}, {Chen}, {Chen},
  {Christlieb}, {Chu}, {Cui}, {Dong}, {Du}, {Fan}, {Feng}, {Fu}, {Gao}, {Gong},
  {Gu}, {Guo}, {Han}, {He}, {Hou}, {Hou}, {Hou}, {Hu}, {Hu}, {Hu}, {Huo},
  {Jia}, {Jiang}, {Jiang}, {Jiang}, {Jin}, {Kong}, {Kong}, {Lei}, {Li}, {Li},
  {Li}, {Li}, {Li}, {Li}, {Li}, {Li}, {Li}, {Li}, {Li}, {Li}, {Liang}, {Lin},
  {Liu}, {Liu}, {Liu}, {Liu}, {Lu}, {Luo}, {Mao}, {Newberg}, {Ni}, {Qi}, {Qi},
  {Shen}, {Shi}, {Song}, {Song}, {Su}, {Su}, {Tang}, {Tao}, {Tian}, {Wang},
  {Wang}, {Wang}, {Wang}, {Wang}, {Wang}, {Wang}, {Wang}, {Wang}, {Wang},
  {Wang}, {Wang}, {Wang}, {Wang}, {Wang}, {Wang}, {Wang}, {Wang}, {Wang},
  {Wang}, {Wei}, {Wei}, {Wu}, {Wu}, {Wu}, {Wu}, {Xing}, {Xu}, {Xu}, {Xu},
  {Yan}, {Yang}, {Yang}, {Yang}, {Yang}, {Yao}, {Yu}, {Yuan}, {Yuan}, {Yuan},
  {Yuan}, {Zhai}, {Zhang}, {Zhang}, {Zhang}, {Zhang}, {Zhang}, {Zhang},
  {Zhang}, {Zhang}, {Zhao}, {Zhou}, {Zhou}, {Zhu}, {Zhu}, {Zou}, \&
  {Zuo}}]{LMTDR8}
{Luo}, A.~L., {Zhao}, Y.-H., {Zhao}, G., {et~al.} 2015, Research in Astronomy
  and Astrophysics, 15, 1095, \dodoi{10.1088/1674-4527/15/8/002}

\bibitem[{{Lutz} {et~al.}(1996){Lutz}, {Feuchtgruber}, {Genzel}, {Kunze},
  {Rigopoulou}, {Spoon}, {Wright}, {Egami}, {Katterloher}, {Sturm},
  {Wieprecht}, {Sternberg}, {Moorwood}, \& {de Graauw}}]{Lutz96_LawGC}
{Lutz}, D., {Feuchtgruber}, H., {Genzel}, R., {et~al.} 1996, \aap, 315, L269

\bibitem[{{Megeath} {et~al.}(2012){Megeath}, {Gutermuth}, {Muzerolle},
  {Kryukova}, {Flaherty}, {Hora}, {Allen}, {Hartmann}, {Myers}, {Pipher},
  {Stauffer}, {Young}, \& {Fazio}}]{OrionSurvey}
{Megeath}, S.~T., {Gutermuth}, R., {Muzerolle}, J., {et~al.} 2012, \aj, 144,
  192, \dodoi{10.1088/0004-6256/144/6/192}

\bibitem[{{Meingast} {et~al.}(2018){Meingast}, {Alves}, \&
  {Lombardi}}]{Meingast2018_OrionLaw}
{Meingast}, S., {Alves}, J., \& {Lombardi}, M. 2018, \aap, 614, A65,
  \dodoi{10.1051/0004-6361/201731396}

\bibitem[{{Page} {et~al.}(2012){Page}, {Brindle}, {Talavera}, {Still}, {Rosen},
  {Yershov}, {Ziaeepour}, {Mason}, {Cropper}, {Breeveld}, {Loiseau}, {Mignani},
  {Smith}, \& {Murdin}}]{XMM}
{Page}, M.~J., {Brindle}, C., {Talavera}, A., {et~al.} 2012, \mnras, 426, 903,
  \dodoi{10.1111/j.1365-2966.2012.21706.x}

\bibitem[{{Rebull} {et~al.}(2010){Rebull}, {Padgett}, {McCabe}, {Hillenbrand},
  {Stapelfeldt}, {Noriega-Crespo}, {Carey}, {Brooke}, {Huard}, {Terebey},
  {Audard}, {Monin}, {Fukagawa}, {G{\"u}del}, {Knapp}, {Menard}, {Allen},
  {Angione}, {Baldovin-Saavedra}, {Bouvier}, {Briggs}, {Dougados}, {Evans},
  {Flagey}, {Guieu}, {Grosso}, {Glauser}, {Harvey}, {Hines}, {Latter},
  {Skinner}, {Strom}, {Tromp}, \& {Wolf}}]{TaurusSurvey}
{Rebull}, L.~M., {Padgett}, D.~L., {McCabe}, C.~E., {et~al.} 2010, \apjs, 186,
  259, \dodoi{10.1088/0067-0049/186/2/259}

\bibitem[{{Rieke} {et~al.}(2004){Rieke}, {Young}, {Engelbracht}, {Kelly},
  {Low}, {Haller}, {Beeman}, {Gordon}, {Stansberry}, {Misselt}, {Cadien},
  {Morrison}, {Rivlis}, {Latter}, {Noriega-Crespo}, {Padgett}, {Stapelfeldt},
  {Hines}, {Egami}, {Muzerolle}, {Alonso-Herrero}, {Blaylock}, {Dole}, {Hinz},
  {Le Floc'h}, {Papovich}, {P{\'e}rez-Gonz{\'a}lez}, {Smith}, {Su}, {Bennett},
  {Frayer}, {Henderson}, {Lu}, {Masci}, {Pesenson}, {Rebull}, {Rho}, {Keene},
  {Stolovy}, {Wachter}, {Wheaton}, {Werner}, \& {Richards}}]{MIPS}
{Rieke}, G.~H., {Young}, E.~T., {Engelbracht}, C.~W., {et~al.} 2004, \apjs,
  154, 25, \dodoi{10.1086/422717}

\bibitem[{{Schlafly} {et~al.}(2016){Schlafly}, {Meisner}, {Stutz},
  {Kainulainen}, {Peek}, {Tchernyshyov}, {Rix}, {Finkbeiner}, {Covey}, {Green},
  {Bell}, {Burgett}, {Chambers}, {Draper}, {Flewelling}, {Hodapp}, {Kaiser},
  {Magnier}, {Martin}, {Metcalfe}, {Wainscoat}, \& {Waters}}]{Schlafly2016_Rv}
{Schlafly}, E.~F., {Meisner}, A.~M., {Stutz}, A.~M., {et~al.} 2016, \apj, 821,
  78, \dodoi{10.3847/0004-637X/821/2/78}

\bibitem[{{Skrutskie} {et~al.}(2006){Skrutskie}, {Cutri}, {Stiening},
  {Weinberg}, {Schneider}, {Carpenter}, {Beichman}, {Capps}, {Chester},
  {Elias}, {Huchra}, {Liebert}, {Lonsdale}, {Monet}, {Price}, {Seitzer},
  {Jarrett}, {Kirkpatrick}, {Gizis}, {Howard}, {Evans}, {Fowler}, {Fullmer},
  {Hurt}, {Light}, {Kopan}, {Marsh}, {McCallon}, {Tam}, {Van Dyk}, \&
  {Wheelock}}]{2MASS}
{Skrutskie}, M.~F., {Cutri}, R.~M., {Stiening}, R., {et~al.} 2006, \aj, 131,
  1163, \dodoi{10.1086/498708}

\bibitem[{{Sun} {et~al.}(2021){Sun}, {Jiang}, {Yuan}, \& {Li}}]{Sun2021_UVmap}
{Sun}, M., {Jiang}, B., {Yuan}, H., \& {Li}, J. 2021, \apjs, 254, 38,
  \dodoi{10.3847/1538-4365/abf929}

\bibitem[{{Sun} {et~al.}(2018){Sun}, {Jiang}, {Zhao}, {Gao}, {Gao}, {Jian}, \&
  {Yuan}}]{Sun18_UV_extinction}
{Sun}, M., {Jiang}, B.~W., {Zhao}, H., {et~al.} 2018, \apj, 861, 153,
  \dodoi{10.3847/1538-4357/aac776}

\bibitem[{Virtanen {et~al.}(2020)Virtanen, Gommers, Oliphant, Haberland, Reddy,
  Cournapeau, Burovski, Peterson, Weckesser, Bright, {van der Walt}, Brett,
  Wilson, Millman, Mayorov, Nelson, Jones, Kern, Larson, Carey, Polat, Feng,
  Moore, {VanderPlas}, Laxalde, Perktold, Cimrman, Henriksen, Quintero, Harris,
  Archibald, Ribeiro, Pedregosa, {van Mulbregt}, \& {SciPy 1.0
  Contributors}}]{Scipy}
Virtanen, P., Gommers, R., Oliphant, T.~E., {et~al.} 2020, Nature Methods, 17,
  261, \dodoi{10.1038/s41592-019-0686-2}

\bibitem[{{Wang} \& {Chen}(2019)}]{Wang19_law}
{Wang}, S., \& {Chen}, X. 2019, \apj, 877, 116,
  \dodoi{10.3847/1538-4357/ab1c61}

\bibitem[{{Wang} \& {Chen}(2024)}]{wang24_jwstIR}
---. 2024, \apjl, 964, L3, \dodoi{10.3847/2041-8213/ad2e98}

\bibitem[{{Wang} {et~al.}(2013){Wang}, {Gao}, {Jiang}, {Li}, \&
  {Chen}}]{Wang2013_coalsack}
{Wang}, S., {Gao}, J., {Jiang}, B.~W., {Li}, A., \& {Chen}, Y. 2013, \apj, 773,
  30, \dodoi{10.1088/0004-637X/773/1/30}

\bibitem[{{Wang} \& {Jiang}(2014)}]{W2014_IRlaw}
{Wang}, S., \& {Jiang}, B.~W. 2014, \apjl, 788, L12,
  \dodoi{10.1088/2041-8205/788/1/L12}

\bibitem[{{Weingartner} \& {Draine}(2001)}]{WD01}
{Weingartner}, J.~C., \& {Draine}, B.~T. 2001, \apj, 548, 296,
  \dodoi{10.1086/318651}

\bibitem[{{Whittet} {et~al.}(2013){Whittet}, {Poteet}, {Chiar}, {Pagani},
  {Bajaj}, {Horne}, {Shenoy}, \& {Adamson}}]{Whittet13_L183}
{Whittet}, D.~C.~B., {Poteet}, C.~A., {Chiar}, J.~E., {et~al.} 2013, \apj, 774,
  102, \dodoi{10.1088/0004-637X/774/2/102}

\bibitem[{{Xue} {et~al.}(2016){Xue}, {Jiang}, {Gao}, {Liu}, {Wang}, \&
  {Li}}]{Xue2016_IRlaw}
{Xue}, M., {Jiang}, B.~W., {Gao}, J., {et~al.} 2016, \apjs, 224, 23,
  \dodoi{10.3847/0067-0049/224/2/23}

\bibitem[{{Yershov}(2014)}]{UVOT}
{Yershov}, V.~N. 2014, \apss, 354, 97, \dodoi{10.1007/s10509-014-1944-5}

\bibitem[{{York} {et~al.}(2000){York}, {Adelman}, {Anderson}, {Anderson},
  {Annis}, {Bahcall}, {Bakken}, {Barkhouser}, {Bastian}, {Berman}, {Boroski},
  {Bracker}, {Briegel}, {Briggs}, {Brinkmann}, {Brunner}, {Burles}, {Carey},
  {Carr}, {Castander}, {Chen}, {Colestock}, {Connolly}, {Crocker}, {Csabai},
  {Czarapata}, {Davis}, {Doi}, {Dombeck}, {Eisenstein}, {Ellman}, {Elms},
  {Evans}, {Fan}, {Federwitz}, {Fiscelli}, {Friedman}, {Frieman}, {Fukugita},
  {Gillespie}, {Gunn}, {Gurbani}, {de Haas}, {Haldeman}, {Harris}, {Hayes},
  {Heckman}, {Hennessy}, {Hindsley}, {Holm}, {Holmgren}, {Huang}, {Hull},
  {Husby}, {Ichikawa}, {Ichikawa}, {Ivezi{\'c}}, {Kent}, {Kim}, {Kinney},
  {Klaene}, {Kleinman}, {Kleinman}, {Knapp}, {Korienek}, {Kron}, {Kunszt},
  {Lamb}, {Lee}, {Leger}, {Limmongkol}, {Lindenmeyer}, {Long}, {Loomis},
  {Loveday}, {Lucinio}, {Lupton}, {MacKinnon}, {Mannery}, {Mantsch}, {Margon},
  {McGehee}, {McKay}, {Meiksin}, {Merelli}, {Monet}, {Munn}, {Narayanan},
  {Nash}, {Neilsen}, {Neswold}, {Newberg}, {Nichol}, {Nicinski}, {Nonino},
  {Okada}, {Okamura}, {Ostriker}, {Owen}, {Pauls}, {Peoples}, {Peterson},
  {Petravick}, {Pier}, {Pope}, {Pordes}, {Prosapio}, {Rechenmacher}, {Quinn},
  {Richards}, {Richmond}, {Rivetta}, {Rockosi}, {Ruthmansdorfer}, {Sandford},
  {Schlegel}, {Schneider}, {Sekiguchi}, {Sergey}, {Shimasaku}, {Siegmund},
  {Smee}, {Smith}, {Snedden}, {Stone}, {Stoughton}, {Strauss}, {Stubbs},
  {SubbaRao}, {Szalay}, {Szapudi}, {Szokoly}, {Thakar}, {Tremonti}, {Tucker},
  {Uomoto}, {Vanden Berk}, {Vogeley}, {Waddell}, {Wang}, {Watanabe},
  {Weinberg}, {Yanny}, {Yasuda}, \& {SDSS Collaboration}}]{SDSS16_maglim}
{York}, D.~G., {Adelman}, J., {Anderson}, John~E., J., {et~al.} 2000, \aj, 120,
  1579, \dodoi{10.1086/301513}

\bibitem[{{Zhang} \& {Yuan}(2023)}]{Zhang2023_CERs}
{Zhang}, R., \& {Yuan}, H. 2023, \apjs, 264, 14,
  \dodoi{10.3847/1538-4365/ac9dfa}

\bibitem[{{Zhang} {et~al.}(2023){Zhang}, {Yuan}, \& {Chen}}]{Zhang2023_RvMap}
{Zhang}, R., {Yuan}, H., \& {Chen}, B. 2023, \apjs, 269, 6,
  \dodoi{10.3847/1538-4365/acf764}

\end{thebibliography}

\begin{table}
\caption{The information of the photometric catalog used}\label{table:sources}
\centering
\begin{threeparttable}
\begin{tabular}{cccccc}
\toprule
\multirow{2}{*}{Survey} & \multirow{2}{*}{Band} & \multicolumn{2}{c}{Data Selection}  & \multicolumn{2}{c}{blue-edge Parameter} \\
\cmidrule(lr){3-4}\cmidrule(lr){5-6}
              &            & $\ab{\lambda}{eff}(\mu m)$\tnote{(a)}   & Numbers\tnote{(b)} &Color  & $\sigma_{\lambda_{1}\lambda_{2}}$\tnote{(c)} \\
\hline\hline
\multirow{2}{*}{GALEX}     & FUV          & $0.153$    & 30k     &FUV-$\Gbp$              & $0.099$ \\
                           & NUV          & $0.231$    & 118k    &NUV-$\Gbp$              & $0.079$ \\
\hline
\multirow{3}{*}{OM}        & UVW2         & $0.212$    & 3.9k    &$\ab{UVW2}{OM}-\Gbp$    & $0.095$ \\
                           & UVM2         & $0.231$    & 10.4k   &$\ab{UVM2}{OM}-\Gbp$    & $0.077$ \\
                           & UVW1         & $0.291$    & 18.6k   &$\ab{UVW1}{OM}-\Gbp$    & $0.044$ \\
\hline
\multirow{3}{*}{UVOT}      & UVW2         & $0.209$    & 27.0k   &$\ab{UVW2}{UVOT}-\Gbp$  & $0.064$ \\
                           & UVM2         & $0.225$    & 20.7k   &$\ab{UVM2}{UVOT}-\Gbp$  & $0.072$ \\
                           & UVW1         & $0.268$    & 27.4k   &$\ab{UVW1}{UVOT}-\Gbp$  & $0.053$ \\
\hline
\multirow{2}{*}{Gaia}      & $\ab{G}{BP}$ & $0.532$    & 3.28m   &$\setminus$             & $\setminus$ \\
                           & $\ab{G}{RP}$ & $0.797$    & 3.28m   &$\Gbp-\ab{G}{RP}$       & $0.030$ \\
\hline
\multirow{5}{*}{SDSS}      & $u$          & $0.357$    & 418k    &$\ab{u}{SDSS}-\Gbp$     & $0.034$ \\
                           & $g$          & $0.475$    & 430k    &$\ab{g}{SDSS}-\Gbp$     & $0.031$ \\
                           & $r$          & $0.620$    & 410k    &$\Gbp-\ab{r}{SDSS}$     & $0.031$ \\
                           & $i$          & $0.752$    & 407k    &$\Gbp-\ab{i}{SDSS}$     & $0.031$ \\
                           & $z$          & $0.899$    & 434k    &$\Gbp-\ab{z}{SDSS}$     & $0.031$ \\
\hline
\multirow{5}{*}{PS1}       & $g$          & $0.490$    & 2.16m   &$\ab{g}{PS1}-\Gbp$      & $0.031$ \\
                           & $r$          & $0.624$    & 1.32m   &$\Gbp-\ab{r}{PS1}$      & $0.031$ \\
                           & $i$          & $0.756$    & 1.15m   &$\Gbp-\ab{i}{PS1}$      & $0.031$ \\
                           & $z$          & $0.869$    & 1.93m   &$\Gbp-\ab{z}{PS1}$      & $0.031$ \\
                           & $y$          & $0.964$    & 2.87m   &$\Gbp-\ab{y}{PS1}$      & $0.031$ \\
\hline
\multirow{3}{*}{2MASS}     & $J$          & $1.24$     & 3.70m   &$\Gbp-J$                & $0.040$ \\
                           & $H$          & $1.66$     & 3.60m   &$\Gbp-H$                & $0.042$ \\
                           & $\K$         & $2.16$     & 3.43m   &$\Gbp-\K$               & $0.041$ \\
\hline
\multirow{2}{*}{WISE}      & W1           & $3.35$     & 3.75m   &$\Gbp-W1$               & $0.040$ \\
                           & W2           & $4.60$     & 3.63m   &$\Gbp-W2$               & $0.041$ \\
\hline\hline
\multirow{3}{*}{2MASS}     & $J$          & $1.24$     & 361k    &$J-\K$                  & $0.045$ \\
                           & $H$          & $1.66$     & 361k    &$H-\K$                  & $0.046$ \\
                           & $\K$         & $2.16$     & 360k    &$\setminus$             & $\setminus$ \\
\hline
\multirow{4}{*}{WISE}      & W1           & $3.35$     & 347k    &$\K-W1$                 & $0.045$ \\
                           & W2           & $4.60$     & 348k    &$\K-W2$                 & $0.044$ \\
                           & W3           & $11.56$    & 207k    &$\K-W3$                 & $0.059$ \\
                           & W4           & $22.09$    & 16k     &$\K-W4$                 & $0.081$ \\
\hline
\multirow{5}{*}{Spitzer}   & $[3.6]$      & $3.6$      & 36k     &$\K-[3.6]$              & $0.062$ \\
                           & $[4.5]$      & $4.5$      & 36k     &$\K-[4.5]$              & $0.070$ \\
                           & $[5.8]$      & $5.8$      & 36k     &$\K-[5.8]$              & $0.071$ \\
                           & $[8.0]$      & $8.0$      & 36k     &$\K-[8.0]$              & $0.067$ \\
                           & $[24]$       & $24$       & 4.4k    &$\K-[24]$               & $0.099$ \\
\bottomrule
\end{tabular}
\begin{tablenotes}
\item {(a)}: Effective wavelength depends on stellar SED, filter transmission curve and extinction, it is a constant calculated by a  Vega spectrum. The details can be found in Section \ref{sec:wls_shift}.
\item {(b)}: Number of sources after quality control. The detailed quality control and data processing are in Section \ref{sec:data}.
\item {(c)}: Mean error of intrinsic color index.
\end{tablenotes}
\end{threeparttable}
\end{table}

\begin{table}
\caption{Color excess ratio (CER) and its dispersion for the seven sub-clouds}\label{table:curve}
\centering
\begin{tabular}{ll|cc|cc|cc|cc|cc|cc|cc}
\hline\hline
                             &        & \multicolumn{2}{c}{PMCMain} & \multicolumn{2}{c}{OMCA} & \multicolumn{2}{c}{OMCB} & \multicolumn{2}{c}{OMCL} & \multicolumn{2}{c}{TMC1} & \multicolumn{2}{c}{TMC2} & \multicolumn{2}{c}{TMCRing}  \\
                             &        & $CERs$   & $\sigma^{(a)}$           & CERs   & $\sigma$        & CERs   & $\sigma$        & CERs   & $\sigma$        & CERs   & $\sigma$        & CERs   & $\sigma$        & CERs   & $\sigma$            \\
\hline
\multirow{9}{*}{\rotatebox{90}{UV$^{(b)}$}}    & FUV      & -3.429 & 0.345           &        &            &        &           &        &                 & -2.629 & 0.291           & -3.581 & 0.158       & -2.510 & 0.220               \\
                             & NUV                & -2.990 & 0.150           &        &            &        &           & -3.080 & 0.189           & -2.781 & 0.189           & -2.869 & 0.149       & -3.071 & 0.122               \\
                             & $\ab{UVW2}{OM}$    &        &                 &        &            &        &           &        &                 &        &                 &        &             &        &                     \\
                             & $\ab{UVM2}{OM}$    &        &                 &        &            &        &           & -3.376 & 0.123           &        &                 &        &             &        &                     \\
                             & $\ab{UVW1}{OM}$    &        &                 & -1.263 & 0.064      & -1.413 & 0.077     & -1.674 & 0.103           &        &                 & -1.707 & 0.105       &-1.610  & 0.098               \\
                             & $\ab{UVW2}{UVOT}$  & -2.076 & 0.211           &        &            & -2.075 & 0.411     & -1.948 & 0.317           &        &                 &        &             & -2.003 & 0.189               \\
                             & $\ab{UVM2}{UVOT}$  & -3.260 & 0.274           &        &            & -3.522 & 0.372     & -3.669 & 0.220           &        &                 &        &             & -3.257 & 0.290               \\
                             & $\ab{UVW1}{UVOT}$  & -1.657 & 0.137           & -1.734 & 0.148      & -1.867 & 0.195     & -1.845 & 0.132           & -1.942 & 0.201           &        &             & -1.744 & 0.183               \\
                             & $u_{SDSS}$         &        &                 & -1.227 & 0.062      & -1.219 & 0.071     & -1.174 & 0.054           & -1.249 & 0.090           & -1.252 & 0.067       & -1.300 & 0.060               \\
\hline
\multirow{9}{*}{\rotatebox{90}{Optical$^{(b)}$}}  & $g_{SDSS}$     &        &                    & -0.390 & 0.021           & -0.403 & 0.028           & -0.387 & 0.020           & -0.425 & 0.037           & -0.397 & 0.027           & -0.405 & 0.023               \\
                             & $r_{SDSS}$     &        &                    & 0.414  & 0.014           & 0.398  & 0.018           & 0.424  & 0.014           & 0.361  & 0.022           & 0.389  & 0.017           & 0.405  & 0.014               \\
                             & $i_{SDSS}$     &        &                    & 0.927  & 0.017           & 0.909  & 0.017           & 0.924  & 0.012           & 0.876  & 0.035           & 0.899  & 0.015           & 0.915  & 0.013               \\
                             & $z_{SDSS}$     &        &                    & 1.296  & 0.019           & 1.299  & 0.023           & 1.315  & 0.017           & 1.287  & 0.019           & 1.289  & 0.020           & 1.288  & 0.017               \\
                             & $g_{PS1}$   & -0.287 & 0.023              & -0.289 & 0.024           & -0.289 & 0.023           & -0.267 & 0.021           & -0.304 & 0.029           & -0.276 & 0.023           & -0.279 & 0.021               \\
                             & $r_{PS1}$   & 0.398  & 0.018              & 0.418  & 0.014           & 0.409  & 0.015           & 0.434  & 0.015           & 0.381  & 0.018           & 0.409  & 0.015           & 0.413  & 0.014               \\
                             & $i_{PS1}$   & 0.910  & 0.018              & 0.911  & 0.015           & 0.902  & 0.019           & 0.915  & 0.019           & 0.899  & 0.016           & 0.912  & 0.016           & 0.917  & 0.015               \\
                             & $z_{PS1}$   & 1.230  & 0.019              & 1.217  & 0.017           & 1.228  & 0.020           & 1.238  & 0.021           & 1.221  & 0.019           & 1.229  & 0.018           & 1.233  & 0.017               \\
                             & $y_{PS1}$   & 1.438  & 0.020              & 1.436  & 0.018           & 1.442  & 0.020           & 1.453  & 0.020           & 1.436  & 0.022           & 1.445  & 0.019           & 1.414  & 0.018               \\
\hline
\multirow{5}{*}{\rotatebox{90}{IR}} & J    &1.790   &0.036             & 1.771   & 0.033            & 1.807  & 0.032            & 1.809  & 0.033            & 1.797  & 0.043            & 1.786  & 0.032            & 1.773  & 0.032                \\
                             & H           &2.055   &0.050             & 2.011   & 0.041            & 2.048  & 0.040            & 2.067  & 0.041            & 2.064  & 0.061            & 2.053  & 0.044            & 2.037  & 0.041                \\
                             & $\K$        &2.187   &0.055             & 2.138   & 0.045            & 2.172  & 0.047            & 2.201  & 0.043            & 2.201  & 0.068            & 2.183  & 0.049            & 2.159  & 0.044                \\
                             & W1          &2.276   &0.064             & 2.217   & 0.051            & 2.245  & 0.055            & 2.281  & 0.052            & 2.285  & 0.074            & 2.262  & 0.056            & 2.244  & 0.051                \\
                             & W2          &2.320   &0.065             & 2.254   & 0.054            & 2.285  & 0.056            & 2.319  & 0.054            & 2.334  & 0.080            & 2.312  & 0.060            & 2.290  & 0.053                \\
\hline\hline
\multirow{10}{*}{\rotatebox{90}{IR Diffuse}} & H      & 0.348  & 0.022              & 0.328  & 0.025           & 0.352  & 0.023           & 0.333  & 0.022           & 0.347  & 0.021           & 0.341  & 0.023           & 0.350  & 0.021               \\
                             & W1     & -0.193 & 0.027              & -0.197 & 0.026           & -0.193 & 0.025           & -0.183 & 0.025           & -0.182 & 0.026           & -0.171 & 0.024           & -0.143 & 0.022               \\
                             & W2     & -0.318 & 0.027              & -0.324 & 0.030           & -0.310 & 0.027           & -0.267 & 0.026           & -0.327 & 0.026           & -0.304 & 0.025           & -0.272 & 0.024               \\
                             & W3     & -0.305 & 0.083              & -0.324 & 0.101           & -0.331 & 0.084           & -0.222 & 0.075           & -0.286 & 0.055           & -0.260 & 0.050           & -0.301 & 0.056               \\
                             & W4     & -0.522 & 0.093              &        &                 & -0.361 & 0.067           & -0.500 & 0.066           & -0.353 & 0.070           & -0.289 & 0.051           & -0.849 & 0.052               \\
                             & [3.6] & -0.244 & 0.054              & -0.232 & 0.030           & -0.217 & 0.028           &        &                & -0.246 & 0.032        & -0.233 & 0.032           &        &                     \\
                             & [4.5] & -0.326 & 0.054              & -0.307 & 0.034           & -0.298 & 0.030           &        &                 & -0.311 & 0.034        & -0.291 & 0.039           &        &                     \\
                             & [5.8] & -0.374 & 0.041              & -0.336 & 0.038           & -0.338 & 0.033           &        &                 & -0.354 & 0.039        & -0.328 & 0.041           &        &                     \\
                             & [8.0] & -0.364 & 0.043              & -0.331 & 0.036           & -0.326 & 0.031           &        &                 & -0.354 & 0.037        & -0.331 & 0.034           &        &                     \\
                             & [24]  & -0.360 & 0.082              & -0.443 & 0.089           & -0.151 & 0.091           &        &                 & -0.219 & 0.074          & -0.402 & 0.087             &        &                     \\
\hline
\multirow{10}{*}{\rotatebox{90}{IR Dense}}   & H      & 0.365  & 0.026              & 0.366  & 0.023           & 0.354  & 0.022           &        &                 & 0.368  & 0.021           & 0.345  & 0.036           &        &                     \\
                             & W1     & -0.097 & 0.064             & -0.144 & 0.060          & -0.153 & 0.072          &        &                 & -0.106 & 0.035           & -0.151 & 0.052           &        &                     \\
                             & W2     & -0.284 & 0.048             & -0.312 & 0.066          & -0.294 & 0.059          &        &                 & -0.278 & 0.037           & -0.357 & 0.059           &        &                     \\
                             & W3     & -0.328 & 0.123             & -0.129 & 0.249          & -0.239 & 0.210          &        &                 & -0.280 & 0.073           & -0.330 & 0.080           &        &                     \\
                             & W4     &        &                   &        &                &        &                &        &                 &   &           &        &                 &        &                     \\
                             & [3.6] & -0.209 & 0.054              & -0.200 & 0.035          & -0.225 & 0.038           &        &                 & -0.199 & 0.046           & -0.223 & 0.014           &        &                     \\
                             & [4.5] & -0.275 & 0.056              & -0.276 & 0.030          & -0.288 & 0.048           &        &                 & -0.243 & 0.045           & -0.276 & 0.030           &        &                     \\
                             & [5.8] & -0.336 & 0.047              & -0.290 & 0.041          & -0.329 & 0.057           &        &                 & -0.293 & 0.052           & -0.346 & 0.024           &        &                     \\
                             & [8.0] & -0.337 & 0.053              & -0.310 & 0.037          & -0.332 & 0.063           &        &                 & -0.291 & 0.044           & -0.369 & 0.026           &        &                     \\
                             & [24]     & -0.347 & 0.078           & -0.315 & 0.130       & -0.316 & 0.191           &        &                   &  &             &        &                 &        &                     \\
\hline\hline
\end{tabular}
\begin{enumerate}
\item [(a):] $\sigma$ is the dispersion of the residual in the CER fitting.
\item [(b):] The maximum extinction is only $A_{V}\approx2$ in UV band and $A_{V}\approx4$ in optical and IR band, which should be paid attention to.
\item [(c):] Blank indicates no reliable result because the number of stars is fewer than 10.
\end{enumerate}
\end{table}

\begin{table}
\centering
\caption{$R$ values in diffuse regions from UV to NIR}\label{table:R_opt}
\begin{tabular}{l|cc}
\hline\hline
        & ${\rv^{\rm fit}}^{(a)}$  & ${\rbp\ge}^{(b)}$ \\
\hline
TMC1    & 3.094$\pm$0.055   & 2.334$\pm$0.080   \\
TMC2    & 3.250$\pm$0.043   & 2.312$\pm$0.060   \\
TMCRing & 3.163$\pm$0.042   & 2.290$\pm$0.053   \\
PMCMain & 3.173$\pm$0.055   & 2.320$\pm$0.065   \\
OMCA    & 3.380$\pm$0.047   & 2.254$\pm$0.054   \\
OMCB    & 3.515$\pm$0.056   & 2.285$\pm$0.056   \\
OMCL    & 3.445$\pm$0.046   & 2.319$\pm$0.054   \\
\hline
\end{tabular}
\begin{enumerate}
\item [(a):] $\rv$ and $\rv$ are fitted with LAMOST data, which means it only traces the extinction in diffuse region of these molecular clouds.
\item [(b):] $\rbp$ here is the lower limit of true $\rbp$ since we neglect the extinction in W2 band (details in Section \ref{sec:discussion:lamost}).
\end{enumerate}
\end{table}

\begin{table}
\centering
\caption{$R_{\K}$ values in diffuse and dense regions in IR$^{(a)}$}\label{table:R_IR}
\begin{tabular}{l|cccc}
\hline\hline
         & This Work$^{b}$   & WC2019                 & WD01 $\rv=5.5$          & Xue+2016$^{(c)}$ \\
\hline
Diffuse  & 0.341$\pm$0.015   & \multirow{2}{*}{0.324} & \multirow{2}{*}{0.407}  & 0.352$\pm$0.049 \\
Dense    & 0.327$\pm$0.026   &                        &                         & 0.346$\pm$0.054 \\
\hline
\end{tabular}
\begin{enumerate}
\item [(a):] Similar to $\rbp$, the definition of $R_{\K}$ is $A_{\K}/E(J-\K)\approx E(\K-[8.0])/E(J-\K)$, and the value in Table is also the lower limit.
\item [(b):] The value in this column is the average and standard deviation of $E(\K-[8.0])/E(J-\K)$ in these MCs samples, diffuse region means that $\JK<1$ and dense means $\JK>1$.
\item [(c):] Diffuse region in \citet{Xue2016_IRlaw} means that $\JK<0.86$ and dense means $0.86<\JK<1.72$.
\end{enumerate}
\end{table}

\newpage

\end{CJK*}
\end{document}